%% file: paper.tex
\documentclass[twocolumn]{aastex63}
\shorttitle{Streaming instability}
\shortauthors{K. Chen, M-K.\ Lin}

\usepackage{amsmath}
\usepackage{paralist}
\usepackage{url}
\usepackage{bm}
\usepackage{multirow}
\usepackage[T1]{fontenc}
\usepackage[utf8]{inputenc}

\newcommand{\p}{\partial}

\newcommand{\ii}{\mathrm{i}}
\newcommand{\C}{\mathcal{C}^\lambda}

\newcommand{\dd}{\delta}
\newcommand{\real}{\operatorname{Re}}
\newcommand{\imag}{\operatorname{Im}}

\newcommand{\Hdust}{H_\mathrm{d}}
\newcommand{\Hgas}{H_\mathrm{g}}

\newcommand{\rhod}{\rho_\mathrm{d}}
\newcommand{\rhog}{\rho_\mathrm{g}}
\newcommand{\sigd}{\Sigma_\mathrm{d}}
\newcommand{\sigg}{\Sigma_\mathrm{g}}
\newcommand{\tstop}{t_\mathrm{s}}
\newcommand{\tgrow}{t_\mathrm{SI}}
\newcommand{\taus}{\tau_\mathrm{s}}
\newcommand{\st}{\mathrm{St}}
\newcommand{\w}{ \widetilde{W}}
\newcommand{\fdust}{f_\mathrm{d}}
\newcommand{\fg}{f_\mathrm{g}}

\newcommand{\calA}{\mathcal{A}}
\newcommand{\calB}{\mathcal{B}}

\defcitealias{lin12}{L12} 
\defcitealias{lin15}{LY15}
\defcitealias{nelson13}{N13}
\defcitealias{barker15}{BL15}
\defcitealias{mcnally14}{MP14}
\defcitealias{goldreich67}{GS67}
\defcitealias{youdin05a}{YG05}

\def \OmK {\Omega_{\rm K}}

\begin{document}

\title{How efficient is the streaming instability in viscous protoplanetary disks?}
\author{Kan Chen}
\affiliation{Institute of Astronomy, University of Cambridge, Madingley Road, Cambridge CB3 0HA, UK}
\affiliation{Institute of Astronomy and Astrophysics, Academia Sinica, Taipei 10617, Taiwan}
\email{kc571@cam.ac.uk}
\author{Min-Kai Lin}
\affiliation{Institute of Astronomy and Astrophysics, Academia Sinica, Taipei 10617, Taiwan}
\email{mklin@asiaa.sinica.edu.tw}
\correspondingauthor{Min-Kai Lin}

\begin{abstract}
The streaming instability is a popular candidate for planetesimal formation by concentrating dust particles to trigger gravitational collapse. However, its robustness against physical conditions expected in protoplanetary disks is unclear. In particular, particle stirring by turbulence may impede the instability. To quantify this effect, we develop the linear theory of the streaming instability with external turbulence modelled by gas viscosity and particle diffusion. We find the streaming instability is sensitive to turbulence, with growth rates becoming negligible for alpha-viscosity parameters $\alpha \gtrsim \st ^{1.5}$, where $\st$ is the particle Stokes number. We explore the effect of non-linear drag laws, which may be applicable to porous dust particles, and find growth rates are modestly reduced. We also find that gas compressibility increase growth rates by reducing the effect of diffusion. We then apply linear theory to global models of viscous protoplanetary disks. For minimum-mass Solar nebula disk models, we find the streaming instability only grows within disk lifetimes beyond $\sim 10$s of AU, even for cm-sized particles and weak turbulence ($\alpha\sim 10^{-4}$). 
Our results suggest it is rather difficult to trigger the streaming instability in non-laminar protoplanetary disks, especially for small particles.  
\end{abstract}

\section{Introduction}

The size and diversity of the exoplanet population suggest that planet formation is an efficient process. Yet, the formation of planetesimals --- the building blocks of planets --- face several challenges \citep{johansen14}. Dust in protoplanetary disks (PPDs) begin as micron-sized particles, which can grow to mm --- cm sizes via sticking, but growth beyond this size is impeded by bouncing or fragmentation \citep{blum18}. Dust may also be lost due to radial drift as a result of gas drag \citep{whipple72}. 

It is thought that the collective self-gravity of a particle swarm may bypass these barriers by direct gravitational collapse into planetesimals. However, particles must first reach high volume densities relative to the gas for direct collapse \citep{shi13}. This condition may be attained through dust settling, radial drift,  particle traps, or other dust-gas instabilities  \citep{chiang10,johansen14}, such as the `streaming instability' \citep[SI,][hereafter \citetalias{youdin05a}]{youdin05a}. 


The SI is generic phenomenon in rotating disks of dust and gas that can lead to dust clumping \citep{youdin07b,johansen07,bai10,bai10c,kowalik13,yang14}. Although its physical interpretation is subtle \citep{jacquet11, lin17, squire17}, direct numerical simulations show that the SI is effective in triggering the direct gravitational collapse of dust clumps  \citep{johansen09,carrera15, simon16,simon17,schafer17,nesvorny19}, provided that dust particles have reached sufficient size and the local dust-to-gas mass density ratio is of order unity or larger. 

Consequently, the SI is now the de facto mechanism for planetesimal formation and is frequently applied to assess planet formation in complex disk models \citep{drazkowska14,drazkowska16,armitage16,carrera17,ercolano17}. However, the numerical experiments that yield the criteria for the SI are often idealized, which may not fully account for physical conditions expected in real PPDs. An important effect is gas turbulence and particle diffusion \citep{youdin07b}.  

PPDs can host a wide range of hydrodynamic and magnetohydrodynamic (MHD) instabilities that drive turbulence. The magneto-rotational instability is a powerful mechanism to generate turbulence \citep{balbus91}, although in PPDs non-ideal MHD effects weaken it \citep[e.g.][]{lesur14,bai15,simon18}. This gives room for hydrodynamic instabilities to develop, which include the `zombie vortex instability' \citep{marcus15}, `convective overstability' \citep{klahr14}, and `vertical shear instability' \citep{nelson13}. For a recent review of these hydrodynamic instabilities, see \cite{fromang17}, \cite{klahr18}, \cite{lyra19}, and references therein. 



The effect of the resulting turbulence on the SI has not been explored fully. Selected { shearing box} simulations have included the magneto-rotational instability \citep[e.g.][]{balsara09,johansen11,yang18} { or driven turbulence \citep{gole20}. However,} these computationally intensive calculations prohibit a parameter study to evaluate the efficiency of the SI in global PPDs. A first step towards this goal is to apply linear theory to PPD models. This requires modeling the linear SI in turbulent disks. Some effort in this direction has been taken by \cite{auffinger17}, who included a viscous stress tensor to mimic the effects of gas turbulence.

More recently, \cite{umurhan19} extended the original analysis of the SI from \citetalias{youdin05a} to include both gas viscosity and a corresponding particle diffusion. They find the SI is then limited to small range of particle sizes at turbulence strengths expected in PPDs.    

Our ultimate goal in this work is to obtain growth timescales and characteristic lengthscales of the SI in realistic PPDs. This will help us understand the relevance of the SI as a function of radius in PPDs. We take this opportunity to expand upon \cite{umurhan19} by considering compressible gas and exploring non-linear drag laws. We also present complementary calculations using a simplified `one-fluid' model of dusty gas based on \cite{lin17} to verify some results.    


This paper is organized as follows. In \S\ref{basic} we describe the basic, two-fluid framework for studying the linear SI in turbulent disks, including models for gas viscosity and particle diffusion. We list the linearized equations in \S\ref{linear} and first present results from controlled numerical experiments in \S\ref{results}. In \S\ref{application} we apply linear theory to assess the efficiency of the SI in physical PPD models; finding the SI is limited to large radii at tens of AU. We summarize the discuss our results in \S\ref{summary}, including model caveats and future directions. In the Appendix we present a simplified, one-fluid model of dusty gas to explain some of the results found in the full two-fluid treatment.  

\input{model}

\input{linear}

\input{result}

\input{application}
\input{summary}

\acknowledgements
{ We thank the anonymous referee for a prompt and clear report that helped to improve the communication of results.} This work is supported by the Taiwan Ministry of Science and Education (grant 107-2112-M001-043-MY3) and the ASIAA Summer Student Program. We thank O. Umurhan for useful discussions. 

\appendix
\input{appendix}

\bibliographystyle{aasjournal}
\bibliography{ref}

\end{document}

%% file: model.tex
\section{Basic equations and parameters}\label{basic}
We consider a protoplanetary disk comprised of { gas and dust} in orbit
about a central star of mass $M_*$. We use $(\rhog,\, P,\,\bm{V})$ to
denote the density, pressure, and velocity of the gas. 

{ We consider a single species of dust treated as a pressureless fluid  with
density and velocity $(\rhod, \bm{W})$, respectively \citep{jacquet11}}. The two fluids 
interact via drag parameterized by a { single} stopping time $\taus$, which is
prescribed below. { A single-species approximation simplifies the analysis considerably. However, it should be noted that this likely  overestimates the efficiency of the SI, as suggested by recent  generalizations of the SI to multi-species dust in inviscid disks  \citep{krapp19}.}  

We neglect disk self-gravity and magnetic fields. For
simplicity, we also neglect the vertical component of stellar gravity and consider unstratified disks. However, in numerical calculations we will account for stratification when choosing physical parameter values. 

In an inertial frame with cylindrical co-ordinates $(R,\phi,z)$
centered on the star, this two-fluid disk is governed by the
following equations: 

\begin{align}
 &\frac{\p\rhog}{\p t} + \nabla\cdot\left(\rhog\bm{V}\right) =
 0,\label{gas_mass}\\   
  &\frac{\p\rhod}{\p t} + \nabla\cdot\left(\rhod\bm{W}\right)
  =
  \nabla\cdot\left[D\rhog\nabla\left(\frac{\rhod}{\rhog}\right)\right],\label{dust_mass}\\ 
& \frac{\p\bm{V}}{\p t} + \bm{V}\cdot\nabla\bm{V}  = - 
\frac{1}{\rhog}\nabla P - \OmK^2R\hat{\bm{R}} + \frac{1}{\rhog}\nabla\cdot\bm{T} \notag\\
&\phantom{\frac{\p\bm{V}}{\p t} + \bm{V}\cdot\nabla\bm{V}  =} +\frac{\epsilon}{\taus}(\bm{W}-\bm{V}),\label{gas_mom_full}\\ 
&\frac{\p\bm{W}}{\p t} + \bm{W}\cdot\nabla\bm{W}  = -
\OmK^2R\hat{\bm{R}} 
-\frac{1}{\taus}(\bm{W}-\bm{V}),\label{dust_mom}\\
\end{align}
where $D$ is a constant diffusion coefficient  \citep{morfill84};
$\OmK(R) = \sqrt{GM_*/R^3}$ is the Keplerian frequency and $G$ is
the gravitational constant; $\epsilon = \rhod/\rhog$ is the local
dust-to-gas ratio; and $\taus$ is the particle stopping time. We consider isothermal 
gas so that $P=c_s^2\rhog$, where $c_s = \Hgas \OmK$ is a prescribed sound-speed and 
$\Hgas$ is a nominal gas disk thickness.

We assume dust particles are subject to diffusion due to turbulent stirring from the gas. In Eq. \ref{gas_mom_full} we thus include a viscous stress tensor $\bm{T}$ to model gas turbulence:
\begin{align}
\bm{T} = \rhog\nu \left[\nabla\bm{V}+\left(\nabla\bm{V}\right)^\dagger - \frac{2}{3}\bm{I}\nabla\cdot\bm{V}\right], 
\end{align}
where $\nu$ is the kinematic viscosity and $\bm{I}$ is the identity tensor. The last 
term in the momentum equations models dust-gas drag and is described below.

\subsection{Local description}

We consider the local stability of the dusty disk. To do so, we
focus on a small patch of the disk and adopt the shearing box
framework \citep{goldreich65}. The shearing box is centered at a point
$(R_0,\phi_0,0)$ that rotates about the star with angular frequency
$\OmK(R_0)\equiv \Omega_\mathrm{0}$, so $\phi_0 =
\Omega_\mathrm{0}t$. Cartesian co-ordinates ($x,y,z$) in the shearing
box correspond to the $(R,\phi, z)$ directions in the global
disk. Global curvature terms are neglected, as are  
radial gradients in densities and disk temperature.   

In this frame, Keplerian rotation appears as a linear shear flow,
$-q\Omega_0 x\hat{\bm{y}}$ with $q=3/2$. We also define $\bm{w},
\bm{v}$ as the velocity deviations from the background Keplerian shear 
in the rotating, local frame. That is, 
\begin{align} 
\bm{w} = \bm{W} - (R - qx)\Omega_0\hat{\bm{y}},
\end{align}
and similarly for $\bm{v}$. For clarity we drop the sub-script `0' below. 

In terms of velocity fluctuations the two-fluid shearing box equations read 
\begin{align}
&  \frac{\p\rhod}{\p t} + \nabla\cdot\left(\rhod\bm{w}\right) -
  q\Omega x \frac{\p\rhod}{\p y} =  \nabla\cdot\left[D\rhog\nabla\left(\frac{\rhod}{\rhog}\right)\right],\label{dust_eqm}\\
&  \frac{\p\rhog}{\p t} + \nabla\cdot\left(\rhog\bm{v}\right)-
  q\Omega x \frac{\p\rhog}{\p y} = 0,\\  
&  \frac{\p\bm{w}}{\p t} + \bm{w}\cdot\nabla\bm{w} - q\Omega x
  \frac{\p\bm{w}}{\p y}  = 2\Omega w_y \hat{\bm{x}} -
  \frac{\kappa^2}{2\Omega} w_x \hat{\bm{y}} \notag\\
&  \phantom{\frac{\p\bm{w}}{\p t} + \bm{w}\cdot\nabla\bm{w} - q\Omega x
  \frac{\p\bm{w}}{\p y}=} 
-\frac{1}{\taus}(\bm{w}-\bm{v}),\\
&  \frac{\p\bm{v}}{\p t} + \bm{v}\cdot\nabla\bm{v} - q\Omega x
  \frac{\p\bm{v}}{\p y}  = 2\Omega v_y \hat{\bm{x}} -
  \frac{\kappa^2}{2\Omega} v_x \hat{\bm{y}} - c_s^2 \nabla\ln{\rhog}  \notag\\
&  \phantom{
    \frac{\p\bm{v}}{\p t} + \bm{v}\cdot\nabla\bm{v} - q\Omega x
    \frac{\p\bm{v}}{\p y}  =
  }
   + 2\eta \Omega^2 R \hat{\bm{x}} + \frac{1}{\rhog}\nabla\cdot\bm{T} \notag\\ 
&  \phantom{
    \frac{\p\bm{v}}{\p t} + \bm{v}\cdot\nabla\bm{v} - q\Omega x
    \frac{\p\bm{v}}{\p y}  =
  }
+ \frac{\epsilon}{\taus}(\bm{w}-\bm{v}). \label{gas_mom}
\end{align}
The SI only operates in the presence of a global pressure gradient. To
include this effect in a local model, we add a constant forcing in the
the gas  momentum equation (\ref{gas_mom}),
$2\eta\Omega^2R\hat{\bm{x}}$, where  
\begin{align}
  \eta \equiv - \frac{1}{2R\Omega^2}\frac{1}{\rhog}\frac{\p P}{\p R},
\end{align}
is a dimensionless measure of the global pressure gradient \citep{youdin05a}. 
This term causes a relative drift between dust and gas (see
\S\ref{2feqm}) and is essential for the streaming instability.

The basic equations \ref{dust_mass}--\ref{gas_mom} are the same as that in \cite{youdin07b} with the addition of gas viscosity and particle diffusion.

\subsection{Generalized stopping times}\label{generalized_tstop}
The magnitude of dust-gas drag is described by the stopping time $\taus$, which is the characteristic decay timescale for a dust particle's velocity relative to the gas, $\Delta v \equiv \left|\bm{v}-\bm{w}\right|$. Smaller $\taus$ indicates stronger coupling between gas and dust. 

Physically, $\taus$ depends on the particle size $a_p$, its internal density $\rho_\bullet$, its relative drift $\Delta v$, the gas density, and the sound-speed \citep{whipple72, weiden77}. The specific form of $\taus(a_p,\rho_\bullet,\Delta v,\rhog, c_s)$ depends on the particle size relative to the mean free path of gas molecules, $\lambda_\mathrm{mfp}$. Particles with $a_p \lesssim  9\lambda_\mathrm{mfp}/4$ are in the Epstein regime with
\begin{align}\label{epstein}
  \taus^\mathrm{Epstein} = \frac{a_p\rho_\bullet}{c_s\rhog}.  
\end{align}
We remark that most studies of the SI assume an Epstein drag law. 

Particles with $a_p\gtrsim 9\lambda_\mathrm{mfp}/4$ enter the Stokes regime. In this case $\taus$ also depends on the Reynolds number defined by  $\mathrm{Re}  \equiv 2 a \Delta v/\nu_m$, where $\nu_m \equiv (1/2)c_s\lambda_\mathrm{mfp}$ is the gas molecular viscosity:
\begin{align}\label{stokes}
  \taus^\mathrm{Stokes} = \begin{cases}
    \frac{2\rho_\bullet a_p^2}{9\nu_m\rhog} & \mathrm{Re} < 1, \\
    \frac{2^{0.6}\rho_\bullet a_p^{1.6}}{9 \nu_m^{0.6} \rhog^{1.4} \Delta v^{0.4}} & 1 < \mathrm{Re} < 800, \\
    \frac{6\rho_\bullet a_p}{\rhog \Delta v} & \mathrm{Re} > 800 
    \end{cases}
\end{align}
\citep{birnstiel10}. Note that $\lambda_\mathrm{mfp}\propto 1/\rhog$. 

One goal of this work is to examine the effect of non-linear drag laws, i.e. when $\taus$ itself depends on the relative drift. As such, instead of adopting different functional forms of $\taus$ that depends on the physical conditions, we use the following  generalized form of $\taus$: 
\begin{align}
  \taus =  \tau_\mathrm{s,eqm}\frac{\rho_\mathrm{g,eqm}^a\left|\bm{w}-\bm{v}\right|^b_\mathrm{eqm}}{\rhog^a\left|\bm{w}-\bm{v}\right|^b},\label{taus_def} 
\end{align}
where $a,b$ are constant parameters, and subscript `eqm' denote equilibrium values. 

Eq. \ref{taus_def} encapsulates the different drag laws described in Eq. \ref{epstein}---\ref{stokes}. For example, the Epstein regime corresponds to $a = 1, b = 0$ and the fully non-linear Stokes law for $\mathrm{Re}>800$ corresponds to $a=b=1$. We find results are insensitive to the index $a$ since the SI does not require compressible gas \citep{youdin05a}. We thus fix $a=1$ for all calculations presented below. 

For convenience we also define the Stokes number $\st$ as a dimensionless measure of the (equilibrium) stopping time,
\begin{align}
	\st \equiv \tau_\mathrm{s,eqm}\Omega.
\end{align} 
For the most commonly considered case of Epstein drag, $\st$ is equivalent to the particle size for fixed internal densities, since $c_s$ is constant and the SI depends weakly on $\rhog$. 

\subsection{Gas turbulence}
We adapt the standard alpha prescription for modeling gas turbulence \citep{shakura73}: 
\begin{align}
\nu = \alpha c_s \Hgas \left(\frac{\rhog}{\rho_\mathrm{g,eqm}}\right)^\xi, 
\end{align}
where $\alpha$ is the dimensionless viscosity and 
$\xi$ is a free parameter. In practice, we take $\xi=-1$ so that 
the dynamic viscosity $\rhog\nu$ is constant to avoid viscous overstabilities \citep{latter06}. 

\subsection{Dust diffusion}
 
We parameterize particle diffusion via the dimensionless coefficient $\delta$ such that 
\begin{align}
  D = \delta c_s \Hgas, \label{diffusion_coefficient}
\end{align}
and $\delta$ is related to the gas viscosity by 
\begin{align}
\delta = \frac{1 + \st + 4\st^2}{\left(1+\st^2\right)^2}\alpha\label{delta_alpha}
\end{align}
\citep{youdin07,youdin11}. For small particles with $\st\ll 1$ we have $\delta\simeq \alpha$. 


\subsection{Two-fluid equilibrium}\label{2feqm}
The two-fluid shearing box equations (\ref{dust_eqm}---\ref{gas_mom}) admit an axisymmetric, steady state with constant $\rhod$, $\rhog$ and no vertical velocities,  $w_z=v_z=0$. The horizontal velocity fluctuations relative to the Keplerian flow are 
\begin{align}
  &w_x = - \frac{2\st}{\Delta^2}\eta R\OmK,\label{Wx}\\
  &w_y = - \frac{1+\epsilon}{\Delta^2}\eta R\OmK,\\
  &v_x = \frac{2\epsilon\st}{\Delta^2}\eta R\OmK,\\
  &v_y = -\frac{1 + \epsilon + \st^2}{\Delta^2}\eta R\OmK,
\end{align}
where
\begin{align}
\Delta^2 = \st^2 + (1+\epsilon)^2.
\end{align}
For typical disk models with $\eta>0$ (a negative pressure gradient), particles drift inwards while gas is pushed out by the mutual drag force. 

\subsection{Connection with stratified disks}\label{relation_strat}
In an unstratified disk model the equilibrium dust-to-gas ratio $\epsilon$ and dust diffusion coefficient $D$, which is determined by the gas viscosity (Eq. \ref{diffusion_coefficient}---\ref{delta_alpha}), can be set independently. Indeed, we take this approach in our initial calculations. 

Physically, however, an unstratified model represents the disk midplane, and vertical dust settling is balanced by turbulent diffusion  \citep[e.g.][]{fromang06,stoll16,flock17,yang18,lin19}. In this case $\epsilon$ and $D$ are no longer independent. The characteristic dust layer thickness $\Hdust$ can be modelled by 
\begin{align}
  \Hdust = \sqrt{\frac{\delta}{\st + \delta}}\Hgas, \label{hd_hg}
\end{align} 
\citep{dubruelle95,lin19}. The midplane dust-to-gas ratio $\epsilon$ is  given by 
\begin{align} 
  \epsilon = Z\frac{\Hgas}{\Hdust} \label{dg_fixZ} 
\end{align}
\citep{johansen14}, where the local metallicity $Z$ is 
\begin{align}
Z = \frac{\sigd}{\sigg}, 
\end{align}
where $\Sigma_\mathrm{d,g}$ are the surface densities in dust and gas, respectively. In our self-consistent calculations, we determine $\epsilon$ by specifying the metallicity $Z$, Stokes number $\st$, and gas viscosity $\alpha$ (and hence $\delta$).

\subsection{One-fluid models}
In addition to the full, two-fluid treatment of dusty gas described above, 
we also supplement some of our calculations with the `one-fluid' model of dusty gas first described by \cite{laibe14,price15} and further developed by \cite{lin17}. In Appendix \ref{one_fluid_model} we extend the one-fluid model to include dust diffusion and non-linear drag laws and compare it with the full two-fluid treatment. 


%% file: linear.tex
\section{Linear problem}\label{linear}

\subsection{Perturbation equations}

We perturb the above two-fluid system with axisymmetric Eulerian perturbations such that
\begin{align}
  \rhog \to \rhog + \delta\rhog\exp{\left[ \ii\left(k_xx + k_zz\right) +
    \sigma t\right] }
\end{align}
and similarly for other variables, where $k_{x,z}$ are radial and vertical wavenumbers and taken to be positive without loss of generality; and $\sigma$ is the complex frequency with growth rate $s\equiv \real(\sigma)$.  

Dropping the `eqm' sub-scripts for clarity, the linearized equations for the dust fluid read: 
\begin{align}
  \sigma \frac{\dd\rhod}{\rhod} &= - \ii k_xw_x\frac{\dd\rhod}{\rhod} - \ii
  k_x\dd w_x - \ii k_z\dd w_z \notag\\
  &\phantom{=}- D k^2 \left(\frac{\dd\rhod}{\rhod}- \frac{\dd\rhog}{\rhog}\right),\label{lin_dust_mass}\\
  \sigma \dd w_x &= -\ii k_x w_x \dd w_x + 2\Omega \dd w_y \notag\\ & +
  \frac{1}{\taus}\left(w_x -
    v_x\right)\left(\frac{\dd\taus}{\taus}\right) -
  \frac{1}{\taus}\left(\dd w_x -
    \dd v_x\right),\\
  \sigma \dd w_y &= - \ii k_x w_x \dd w_y - \frac{\Omega}{2}\dd w_x \notag\\
  &+\frac{1}{\taus}\left(w_y -
    v_y\right)\left(\frac{\dd\taus}{\taus}\right) -
  \frac{1}{\taus}\left(\dd w_y -
    \dd v_y\right),\\
  \sigma \dd w_z &= -\ii k_x w_x \dd w_z - \frac{1}{\taus}\left(\dd w_z -
  \dd v_z\right),
  \end{align}
and that for the gas equations are:  
  \begin{align}
  \sigma \frac{\dd\rhog}{\rhog} &= - \ii k_xv_x\frac{\dd\rhog}{\rhog} - \ii
  k_x\dd v_x - \ii k_z\dd v_z,\\
\sigma \dd v_x &= -\ii k_xv_x\dd v_x + 2\Omega\dd v_y - \ii k_xc_s^2 
\frac{\dd\rhog}{\rhog} + \delta F^\text{visc}_x \notag\\
& - \frac{\epsilon}{\taus}\left(w_x -
  v_x\right)\left[\frac{\dd\left(\taus\rhog\right)}{\taus\rhog} -
  \frac{\dd\rhod}{\rhod}\right]\notag\\
& + \frac{\epsilon}{\taus}\left(\dd w_x - \dd v_x\right),\\
\sigma \dd v_y &= -\ii k_xv_x\dd v_y - \frac{\Omega}{2}\dd v_x +\delta F^\text{visc}_y\notag\\ 
 &- \frac{\epsilon}{\taus}\left(w_y -
  v_y\right)\left[\frac{\dd\left(\taus\rhog\right)}{\taus\rhog} -
  \frac{\dd\rhod}{\rhod}\right]\notag\\
& + \frac{\epsilon}{\taus}\left(\dd w_y - \dd v_y\right),\\
\sigma \dd v_z &= -\ii k_xv_x\dd v_z - \ii k_zc_s^2 
\frac{\dd\rhog}{\rhog} + \delta F^\text{visc}_z \notag\\
&+ \frac{\epsilon}{\taus}\left(\dd w_z - \dd
  v_z\right)\label{lin_gas_vz}.
\end{align}
In the above equations the linearized viscous forces are 
\begin{align}
\delta F^\text{visc}_x &= -\nu\left(k_z^2 + \frac{4}{3}k_x^2\right)\delta v_x - \frac{1}{3}\nu k_x k_z \delta v_z,\\
\delta F^\text{visc}_y &= -\nu \left(k_z^2 + k_x^2\right)\dd v_y,\\
\delta F^\text{visc}_z & = -\nu\left(\frac{4}{3}k_z^2 + k_x^2\right)\delta v_z - \frac{1}{3}\nu k_x k_z \delta v_x 
\end{align}
\citep{lin16}; and the linearized stopping time is 
\begin{align}
  \frac{\dd\taus}{\taus} =
  -a\frac{\dd\rhog}{\rhog} 
  -\frac{b}{|\bm{w}-\bm{v}|^2}&\left[\left(w_x 
      -v_x\right)\left(\dd w_x-\dd v_x\right)\right. \notag\\ 
  &+ \left.\left(w_y -v_y\right)\left(\dd w_y-\dd v_y\right)\right].  
\end{align}

Eq. \ref{lin_dust_mass}---\ref{lin_gas_vz} constitutes an eigenvalue problem 
\begin{align}
  \bm{M}\bm{q} = \sigma\bm{q},
\end{align}
where $\bm{q}=(\delta \rhod, \delta \bm{w}, \delta \rhod, \delta \bm{v} )^\text{T}$ is the eigenvector and $\bm{M}$ is the matrix representation of the right-hand-side of Eq. \ref{lin_dust_mass}---\ref{lin_gas_vz}. We solve this eigenvalue problem with standard matrix routines provided by the \textsc{lapack} package\footnote{\url{http://www.netlib.org/lapack/}}.


\subsection{Dimensionless parameters}

We solve the stability problem numerically to find the dimensionless SI growth rate $S\equiv  s /\Omega$ as a  function of the following parameters: 
\begin{itemize}
\item $\st$: the Stokes number or particle size. 
\item $\epsilon = \rhod/\rhog$: the equilibrium dust-to-gas ratio. This is either set directly or indirectly via the total metallicity $Z$ (see \S\ref{relation_strat}). 
\item $\alpha$: the gas viscosity parameter, which also determines the particle diffusion strength $\delta$. 
\item $b$: the power-law index that determines the degree of non-linearity in the drag law.  
\item $K_{x,z}\equiv k_{x,z}\eta R$: the dimensionless perturbation wavenumbers. 
\end{itemize}

We also normalize velocities by $\eta R\Omega$. Together with the above normalizations, the linearized equations may be rendered dimensionless, in which case $\eta$ appears as $\eta R\Omega/c_s\equiv \hat{\eta}$ and becomes a measure for gas compressibility \citep{youdin07b}. We set $\hat{\eta} = 0.05$ unless otherwise stated.  
    

%% file: result.tex
\section{Results}\label{results}

\subsection{Epstein drag in inviscid disks}

We begin with a fiducial setup assuming Epstein drag
$(a,b)=(1,0)$ without viscosity or diffusion ($\alpha=\delta=0$). This is the 
standard case considered in previous analytic SI calculations
\citep{youdin05a,kowalik13}. Following \citeauthor{youdin05a}  
we fix $K_z$ ($=30$ here) and maximize growth rates over $K_x$. We
find qualitatively similar behavior for other fixed values of $K_z$. 
For the cases examined below we find the optimum $K_x$ decrease from $O(10^{2-3})$ at $\st=10^{-4}$ to $O(10^{0-1})$ at $\st=1$.

\begin{figure}
\includegraphics[width=\linewidth]{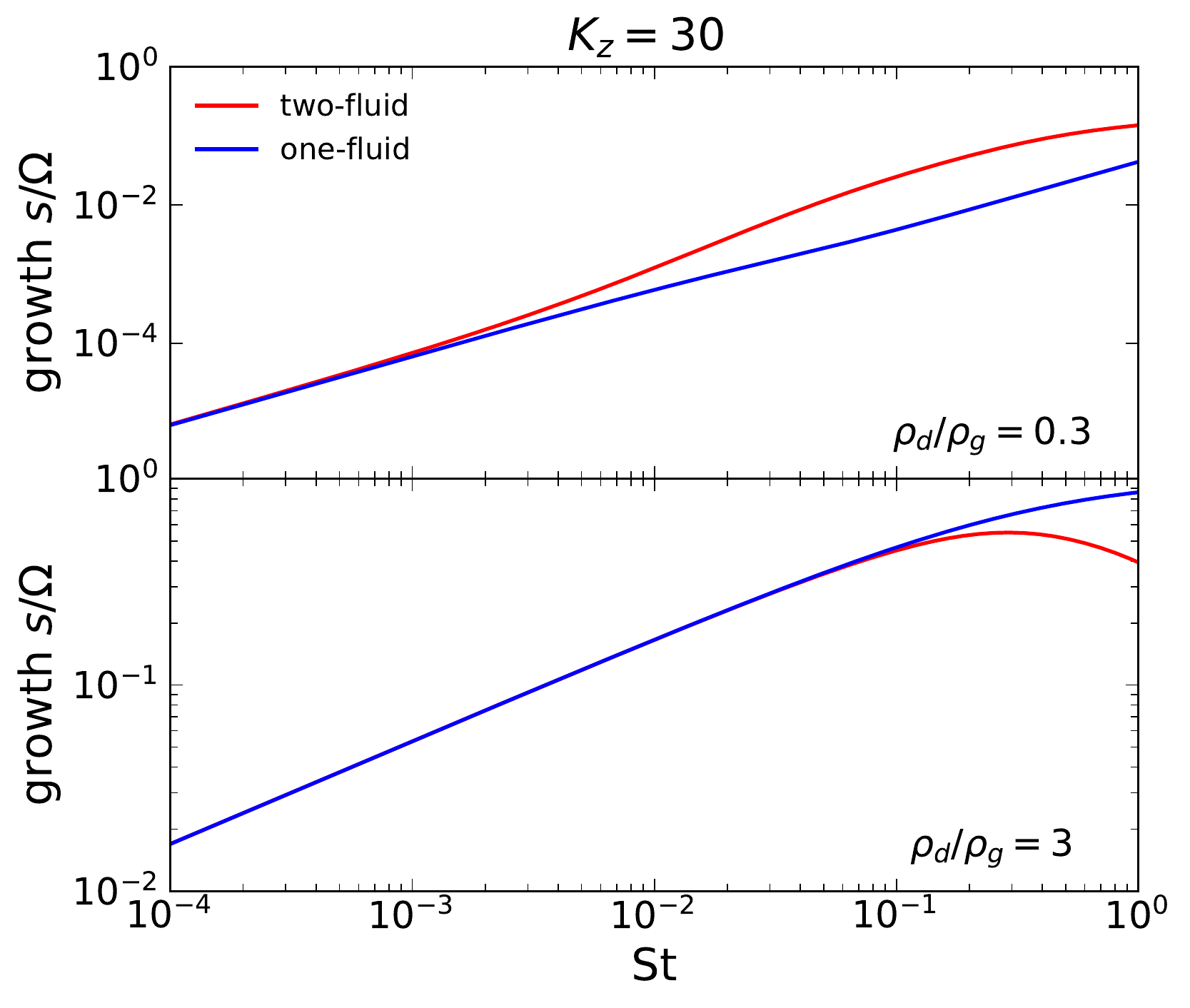}
\caption{Normalized streaming instability growth rates as a function of Stokes number $\st$ in 
  the full two-fluid (red) and simplified one-fluid (blue) models
  using the Epstein drag law  
  ($\tau_s\propto1/\rho_{g}$) without viscosity or diffusion ($\nu=D=0$) 
  The dust-to-gas ratio is $\epsilon=0.3$ (upper) and $\epsilon=3$
  (lower). We fix $K_z=30$ and  
  plot the maximum growth rate over $K_x$. 
\label{fig:41_1}}
\end{figure}

In Fig. \ref{fig:41_1}  we plot growth rates as a function of $\st$ for
two dust-to-gas ratios: a dust-poor disk  with $\epsilon=0.3$ and a
dust-rich disk with $\epsilon=3$. We limit the the Stokes number $\st
< 1$ since larger particles violate the fluid approximation
\citep{jacquet11}. To check our results, we also plot corresponding
growth rates obtained from  the one-fluid framework described in 
Appendix \ref{one_fluid_model}. Our two-fluid results are consistent with earlier
calculations \citep{youdin05a, youdin07b}.

We find the one-fluid approximation is accurate for
$\st\lesssim 0.1$  when $\epsilon > 1$. However, for $\epsilon < 1$,
the one-fluid model only reproduces the full two-fluid results for
$\st \lesssim10^{-3}$. Notice also the one-fluid model tends to over
(under) estimate SI growth rates in dust rich (poor) disks. 


\subsection{Effect of turbulence} 

We now examine turbulent disks by including a gas viscosity $\alpha\neq0$, which determines the particle diffusion coefficient $\delta$ from Eq. \ref{delta_alpha}. We continue with the Epstein drag law with $a=1, b=0$. We either set the dust-to-gas ratio directly as a free parameter, or physically via the total metallicity  $Z=\sigd/\sigg$. We discuss these approaches separately. 

All results in this section are obtained from the full two-fluid equations. In Appendix \ref{analytic_model} we develop a simplified analytic model from the one-fluid approximation, which only includes dust diffusion. 

\subsubsection{Fixed local dust-to-gas ratios}

For fixed $\rhod/\rhog$, turbulence only takes effect through the perturbation equations (cf. fixed metallicity considered later).

\begin{figure}
\includegraphics[width=\linewidth]{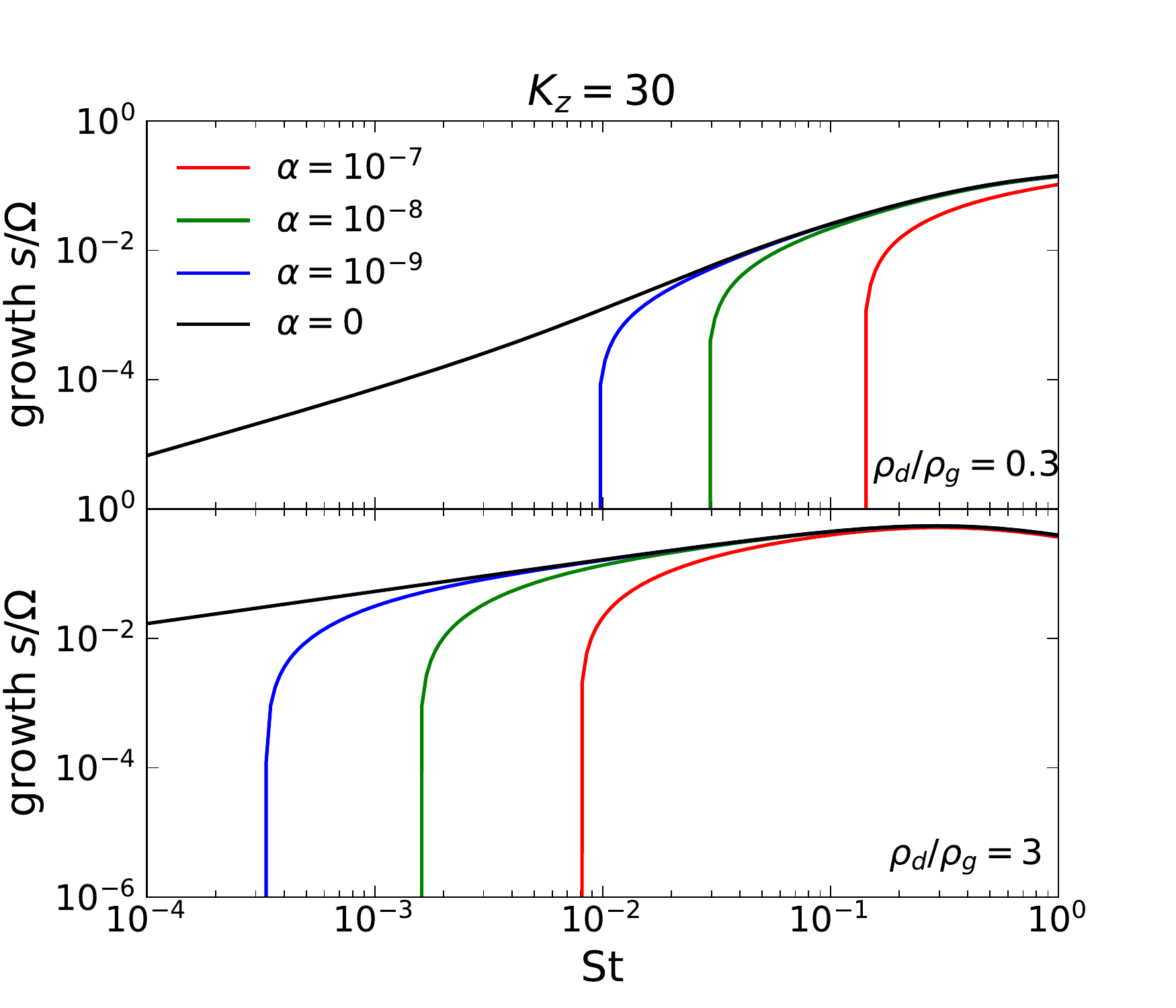}
\caption{Streaming instability growth rates as a function of $\st$ in turbulent disks with particle diffusion. We fix $K_z=30$ and optimize growth rates over $K_x$. Colors of lines denote different values of the gas viscosity parameter $\alpha$. \label{fig:grow_st_fixkz} 
}
\end{figure}

Fig. \ref{fig:grow_st_fixkz} shows growth rates as a function of $\st$ for $\alpha = 0, 10^{-9},\, 10^{-8},\, 10^{-7}$ for fixed $K_z=30$ and maximized over $K_x$. Notice even such small values of $\alpha$ significantly stabilizes the SI. For fixed $K_z$ we find there exists a minimum Stokes number, $\st_\mathrm{min}$, for the SI to exist. $\st_\mathrm{min}$ generally increases with larger $\alpha$ but decreases with increasing $\rhod/\rhog$. This implies that in turbulent disks, larger particles and/or higher dust-to-gas ratios are required for the SI. Equivalently said, small particles are more sensitive to turbulence than larger particles. We find the optimum radial wavenumber decreases from $K_x\sim 10^{3}$ for $\st = 10^{-4}$ to $K_x \sim 1$ for $\st = 1$, but are insensitive to $\alpha$. As shown below, we speculate this is due to viscosity having a larger impact on the SI's vertical structure (which is fixed here) than the radial wavenumber.

\begin{figure*}
\begin{center}
\includegraphics[width=\textwidth]{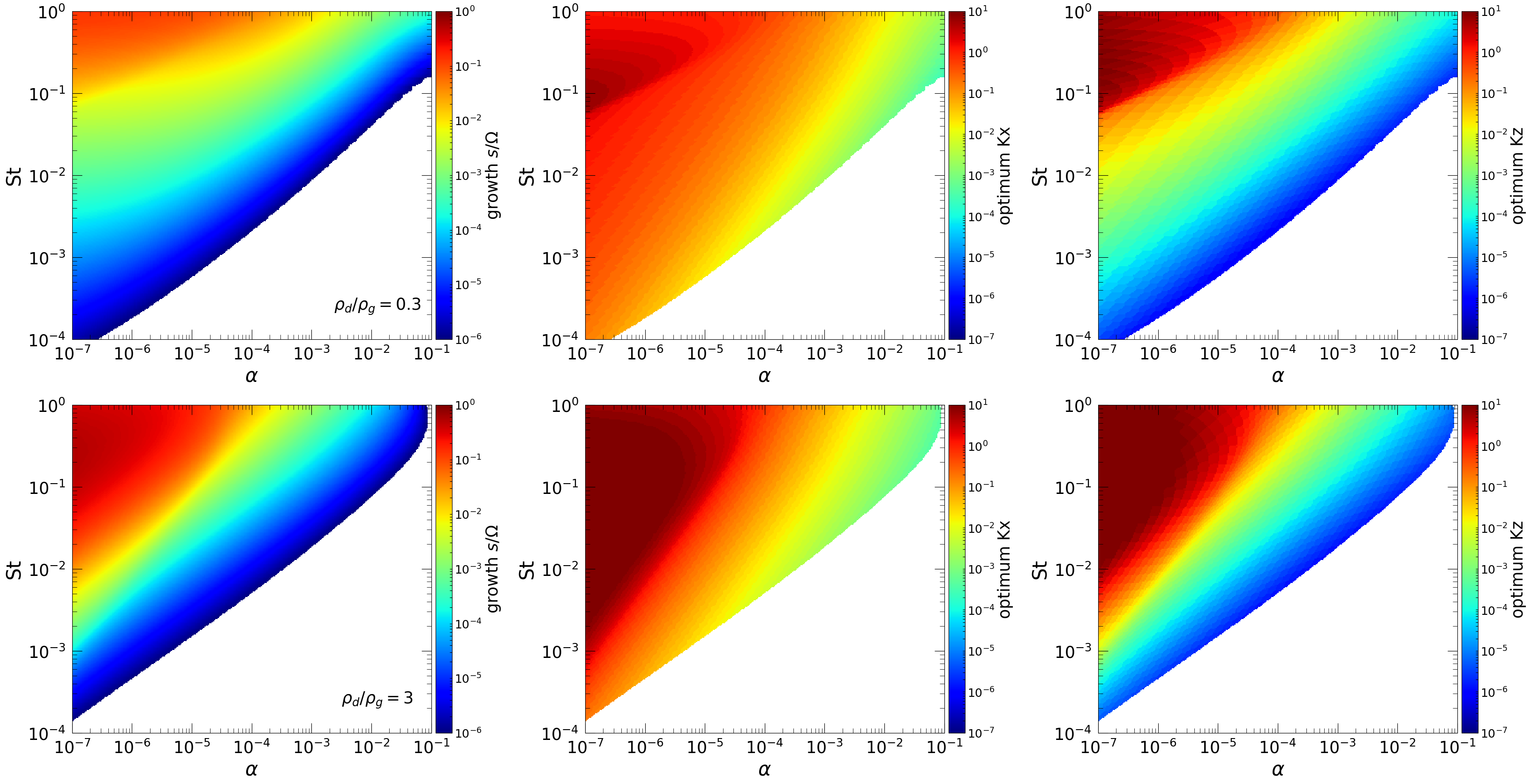}
\end{center}
\caption{Streaming instability growth rates (left) and optimum radial and vertical wavenumbers (middle, right); as a function of gas viscosity with a corresponding dust diffusion coefficient and Stokes number $\st$ for fixed  dust-to-gas ratios $\epsilon=0.3$ (top) and $\epsilon=3$ (bottom). We  truncate the plots for $s <10^{-6}\Omega$. 
\label{fig:con_st_alpha}}
\end{figure*}

Next, we optimize growth rates over $K_z$ as well. Fig. \ref{fig:con_st_alpha} shows the optimum growth rates and wavenumbers for both a dust-poor and dust-rich disk. For a given $\st$, growth rates fall below $10^{-6}\Omega$ when the turbulence exceeds some critical value $\alpha_\mathrm{max}$, which increases with $\st$. However, unlike for  fixed-$K_z$ calculations, here both dust-rich and dust-poor disks have similar values of $\alpha_\mathrm{max}$.  


The middle and right panels of Fig. \ref{fig:con_st_alpha} show the corresponding optimum wavenumbers. As expected, increasing the viscosity generally increase SI lengthscales. Notice also for fixed $\st$ and increasing viscosity that $K_z\ll K_x$, implying that turbulence smears out the SI more easily in the vertical direction. 



\subsubsection{Fixed local metallicities}

We now consider a more physical setup by fixing the total metallicity $Z\equiv \sigd/\sigg$, and setting the dust-to-gas ratio $\epsilon \equiv \rhod/\rhog$ in accordance with dust settling, Eq. \ref{dg_fixZ}. In this case $\rhod/\rhog$ also depends on the Stokes number $\st$ and particle diffusion coefficient $\delta$, which itself is determined by the gas viscosity $\alpha$ (Eq. \ref{delta_alpha}). That is, the basic state now also varies with turbulence strength.


                
\begin{figure*}
\begin{center}
\includegraphics[width=\textwidth]{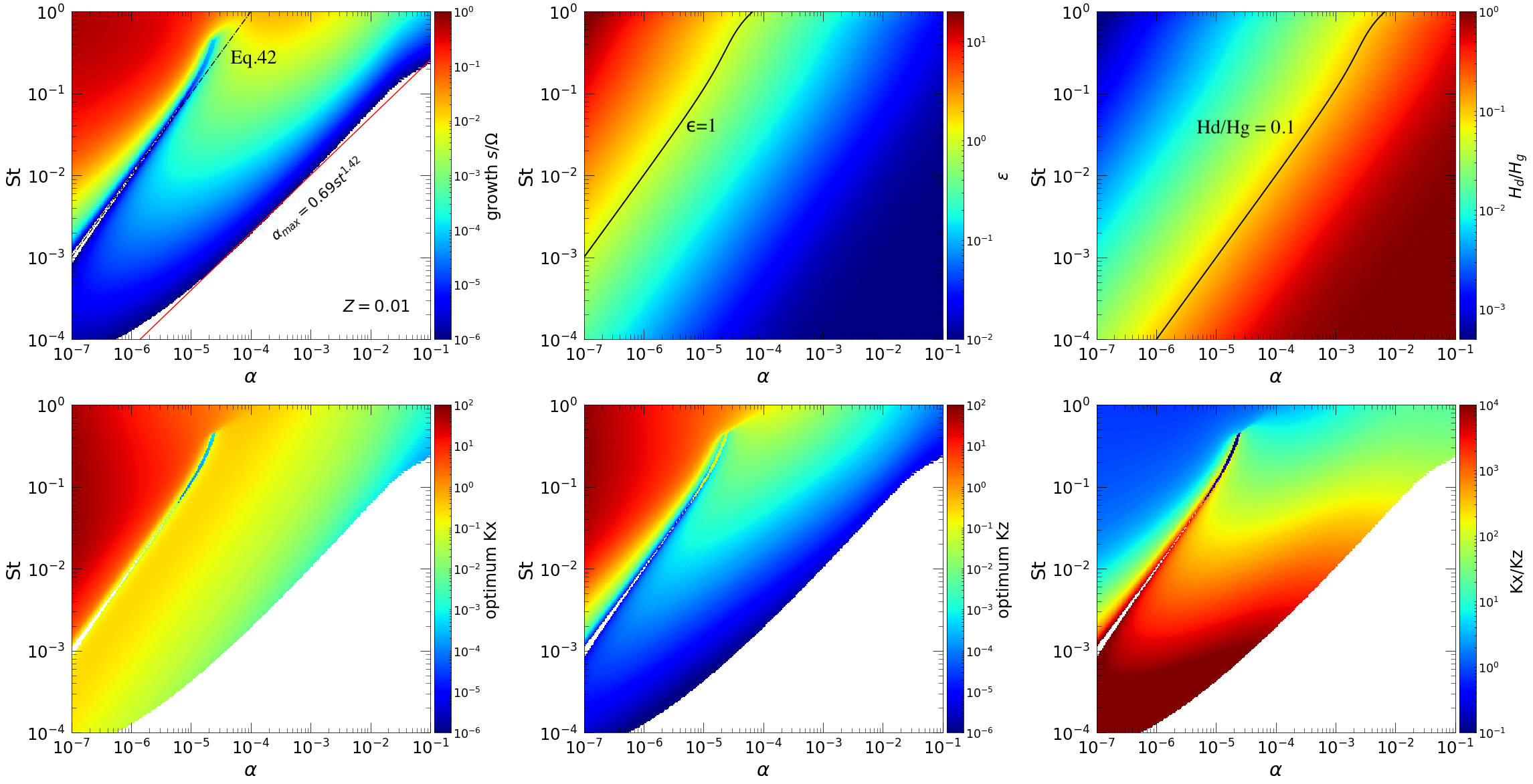}
\end{center}
\caption{Growth rates (upper left) with fixed metallicity $Z=0.01$ as a function of Stokes number and gas viscosity (with a corresponding particle diffusion coefficient).  The red line is an empirical fit to the maximum allowed $\alpha$, beyond which growth rates become negligible  ($<10^{-6}\Omega$). The black dot-dashed line corresponds to Eq. \ref{alpha_st_relation_dg1}. The dust-to-gas ratio (upper middle) and dust scale height normalized by the gas scale height (upper right) are also shown. We also mark some characteristic contours with black lines in these two panels, namely $\epsilon=1$ and $\Hdust/\Hgas=0.1$.   
The corresponding optimum wavenumbers, $K_x$, $K_z$, and their ratio $K_x/K_z$, are shown in the lower left, middle, and right panels,  respectively.
\label{fig:fixZ_0.01}}
\end{figure*}

As a fiducial case we set $Z=0.01$ and plot growth rates and optimum wavenumebrs $K_{x,z}$ in Fig. \ref{fig:fixZ_0.01}, along with $\epsilon$ and $\Hdust/\Hgas$. The dust-to-gas ratio ranges from $\sim10^{-2}$ to $20$; while ${\Hdust}/{\Hgas}$ ranges from $10^{-3}$ to unity. The gap in the upper left of the figure corresponds to $\epsilon \sim 1$ where the SI is quenched \citep{youdin05a}, see also Appendix \ref{analytic_dust_rich}.  

As before, larger $\st$ and smaller $\alpha$ give higher growth rates. The dust-rich SI (left of the gap) involves smaller wavelengths than the dust-poor SI (right of the gap), since viscosity is larger in the latter case. Interestingly, we find $K_x<K_z$ for the dust-rich SI; while $K_x>K_z$ for the dust-poor SI. This implies with high viscosity the SI becomes vertically unstructured. 



SI growth rates are only dynamical ($\sim\Omega$) for $\epsilon > 1$ \citep{youdin05a}. From Eq. \ref{dg_fixZ}, this requires $Z\sqrt{\left(\st+\delta\right)/\delta}>1$. For small $\st$ we can approximate $\delta \simeq \alpha$. Thus the dynamical SI is limited to 
\begin{align}\label{alpha_st_relation_dg1}
\alpha \lesssim \frac{Z^2}{1-Z^2}\st\simeq Z^2\st,  
\end{align} 
where the last equality assumes $Z\ll 1$. It should be noted that here SI is quenched because the background dust-to-gas ratio approaches unity as viscosity is increased from zero; as opposed to the perturbations being  stabilized by viscosity. 

At fixed $\st$, increasing $\alpha$ eventually pushes the system into the dust-poor regime of the SI ($\epsilon < 1$), which is slower than dynamical. Physically this is due to dust becoming vertically mixed by gas turbulence. Further increasing viscosity, the SI is reduced to negligible growth rates. 

\begin{figure*}
\begin{center}
\includegraphics[width=\linewidth]{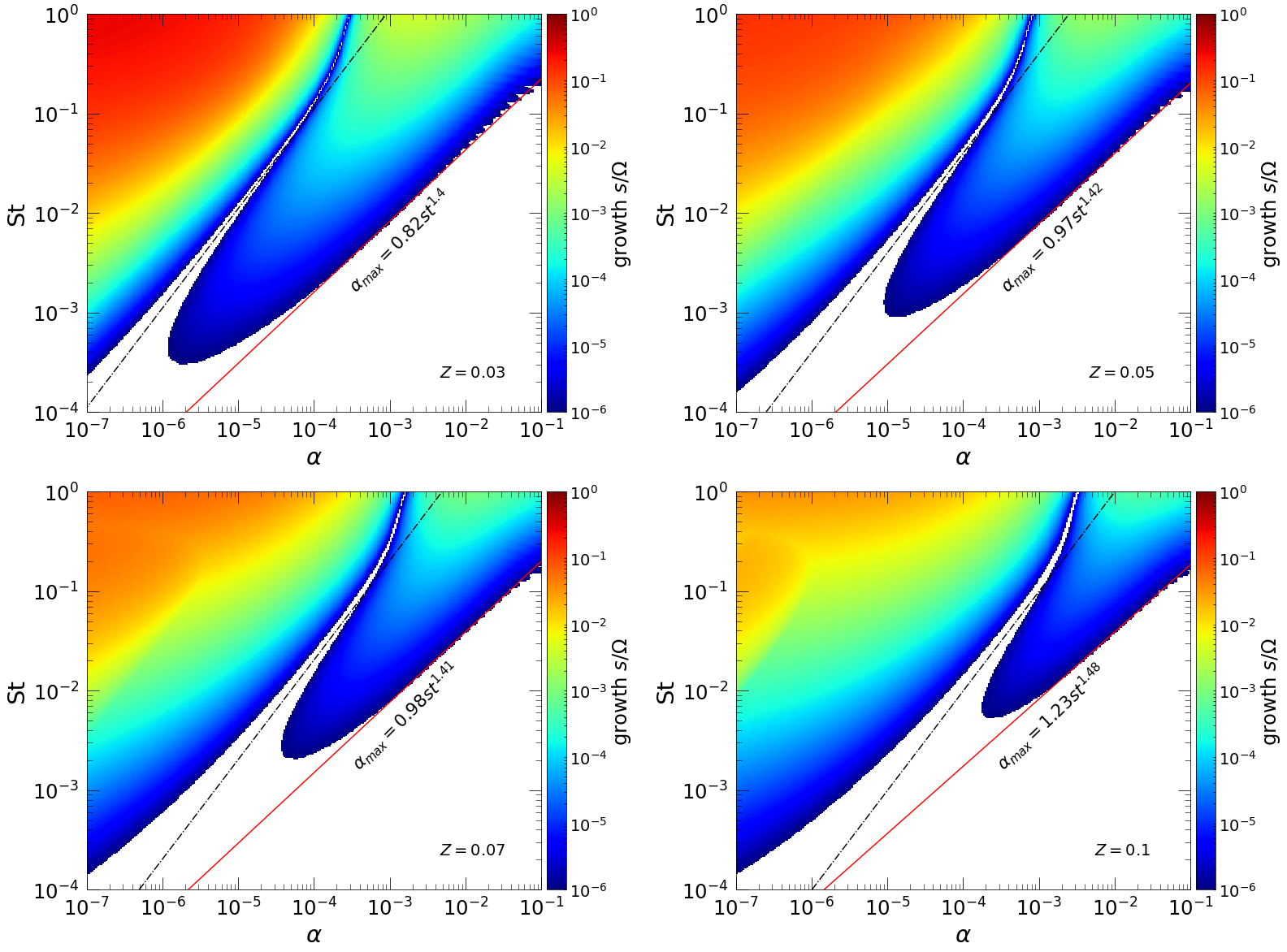}
\caption{Growth rates as a function of $\st$ and $\alpha$ for different metallicities $Z$. The black dot-dashed lines correspond to Eq. \ref{alpha_st_relation_dg1}, while red lines empirical fits to the maximum allowed $\alpha$ for growth rates to remain $>10^{-6}\Omega$.  
\label{fig:fixZ_others}}    
\end{center}
\end{figure*}

Fig. \ref{fig:fixZ_others} show the growth rates as a function of $\st$ and $\alpha$ for other metallicities $Z=0.03, 0.05, 0.07$ and $0.1$. The red line in each panel is a fit to the maximum $\alpha$ as a function of $\st$. These contour plots are essentially translations of that for $Z=0.01$ (to the upper right). As expected, increasing dust loading expands the region of the dust-rich SI. Its boundary becomes less-well approximated by Eq. \ref{alpha_st_relation_dg1}, but it is clear that SI can persist in more turbulent disks if the overall dust content is increased. Notice, however, that the dust-rich SI in fact slows down with increasing $Z$. This is because the SI is also quenched in the limit of gas-free disks \citep{youdin05a}.

We estimate from Fig. \ref{fig:fixZ_0.01}--\ref{fig:fixZ_others} a maximum viscosity $\alpha_\mathrm{max}\sim \st^{1.5}$ above which growth rates become negligible ($<10^{-6}\Omega$). This relation varies weakly with our choice of minimum growth rates or metallicity. Thus, the SI is rapidly quenched by viscosity for small particles.

\subsection{Non-linear drag laws} 

Here we consider the effect of different drag laws on the SI. Recall from \S\ref{generalized_tstop} that our stopping times are parameterized as $\taus \propto \rhog^{-a}\left|\bm{w}-\bm{v}\right|^{-b}$. We find the SI is insensitive to $a$. Thus, we fix $a=1$ and focus on the effect of $b$, i.e the degree of non-linearity. We consider drag laws with $b=0.4$ and $b=1$. 

\subsubsection{Inviscid disks} 
\begin{figure}
 \includegraphics[width=\linewidth]{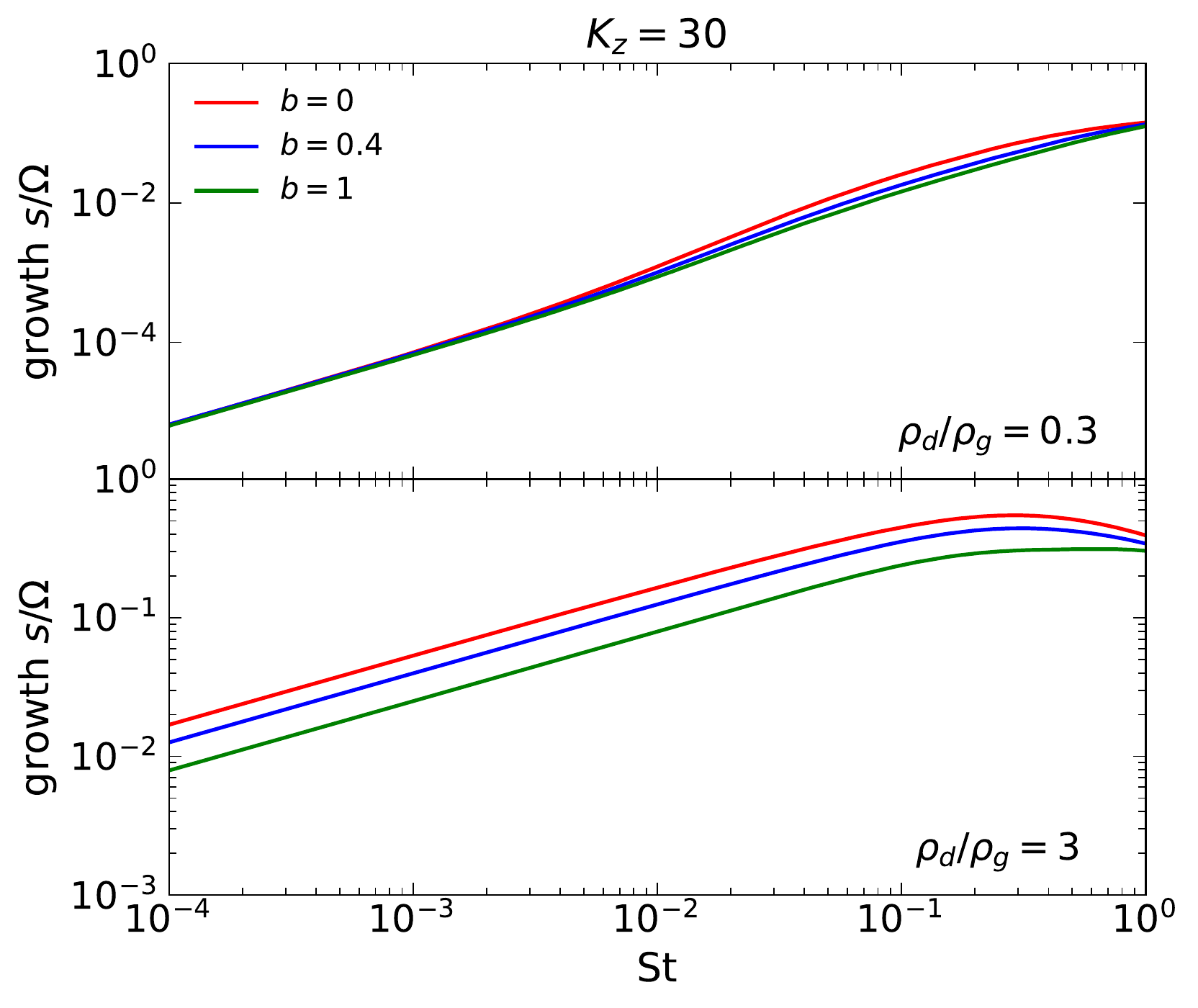}
 \caption{Streaming instability growth rates in inviscid disks as a function of Stokes number 
  for different drag laws with $b=0$ (red, linear drag), $0.4$ (blue, non-linear drag) and $1$ (green, quadratic drag); for $\epsilon=0.3$ (top) and  $\epsilon=3$ (bottom).\label{fig:s_st_b}} 
 \end{figure}

We first return to inviscid disks ($\alpha = \delta = 0$) to isolate the effect of the drag law . Fig. \ref{fig:s_st_b} show growth rates as a function of Stokes number with $\epsilon = 0.3$ and $\epsilon=3$; for $K_z=30$ and maximized over $K_x$. We find the optimum $K_x$ does not vary significantly with $b$. 
 
Increasing the degree of non-linearity reduces SI
growth rates. This effect is small in dust-poor disks, but becomes noticeable in dust-rich disks with 
$\epsilon >1$, although still modest: growth rates are only halved upon increasing $b$ from zero to unity. This is explained in Appendix \ref{analytic_dust_rich} using the one-fluid model of dusty gas.  


\subsubsection{Viscous disks}\label{nonlinear_drag_viscous} 

\begin{figure*}
\begin{center}
\includegraphics[width=\linewidth]{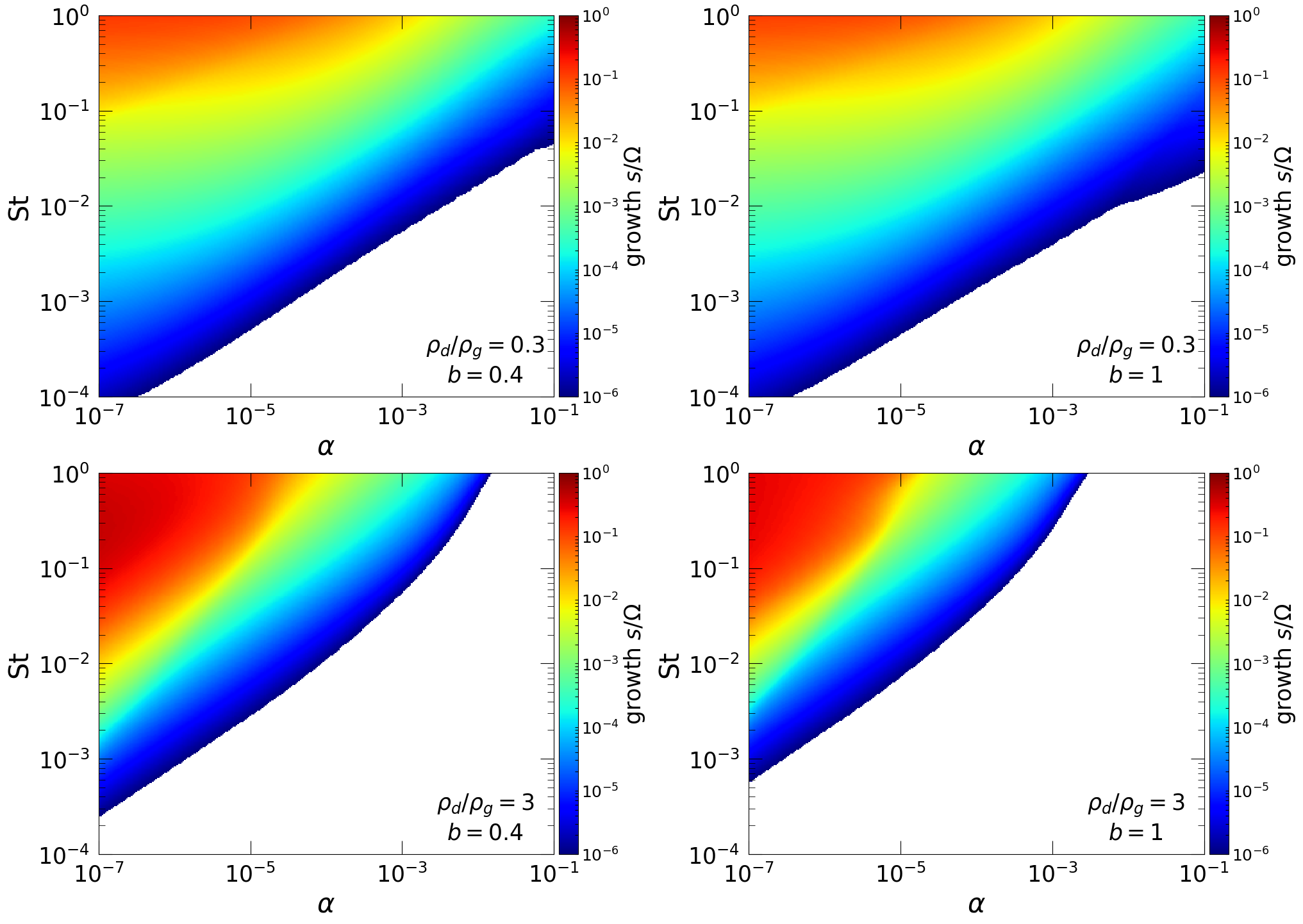}
\end{center}
\caption{Streaming instability growth rates as a function of Stokes number and viscosity for different degrees of non-linearity in the dust-gas drag law, as parameterized by $b$. The top and bottom panels 
show results for $\epsilon =0.3$ and $\epsilon = 3$, respectively.
\label{drag_law_viscous_fix_dg}}
\end{figure*}

We now consider viscous disks and begin with fixing $\rhod/\rhog$ in Fig. \ref{drag_law_viscous_fix_dg}. In the  dust-poor disk with $\rhod  = 0.3\rhog$, increasing $b$ has a negligible effect for $\st\gtrsim 0.1$ and $\st \lesssim 10^{-2}$; while for intermediate $\st$ the SI can persist to slightly higher viscosity with increasing $b$. However, for $\rhod = 3\rhog$, we find increasing $b$ has noticeable effects: larger $b$ makes it easier for viscosity to kill the SI and shrinks the region of instability. 


\begin{figure*}
 \begin{center}
 \includegraphics[width=\linewidth]{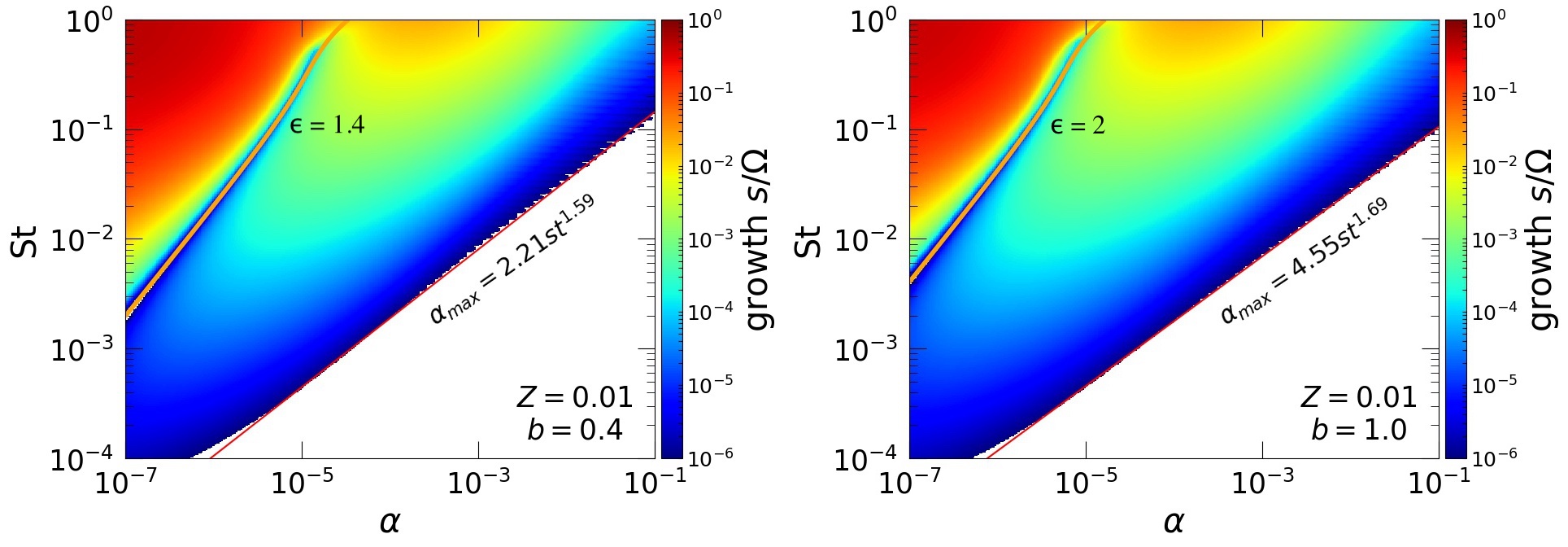}
 \end{center}
 \caption{Similar to Fig. \protect\ref{drag_law_viscous_fix_dg} but for fixed metallicities $Z$. Orange lines denote $\epsilon = 1+b$ where the gaps show. Red lines empirical fits to the maximum allowed $\alpha$ for growth rates to remain $>10^{-6}\Omega$. 
 \label{fig:fixz0.01_a_b}}
 \end{figure*}

We next consider fixing $Z$. In Fig. \ref{fig:fixz0.01_a_b}, we show growth rates with $Z=0.01$ as a function of Stokes number and gas viscosity for different $b$. Notice the growth rate `gap' shifts from $\epsilon = 1$ for $b=0$ (see Fig. \ref{fig:fixZ_0.01}) to $\epsilon =1.4$ for $b=0.4$, and $\epsilon = 2$ for $b=1$. This is in fact consistent with the one-fluid model presented in Appendix \ref{analytic_dust_rich}, for which SI growth rates vanish when $\epsilon = 1 + b$. This explains why the region of `red' SI modes, with dynamical growth rates, shrinks with increasing non-linearity: at fixed $\st$ and increasing $\alpha$ from zero, $\epsilon$ drops to $1+b$ sooner with larger $b$. We find the maximum allowed viscosity shifts from $\alpha_\mathrm{max}\sim \st^{1.5}$ (for $b=0$) to $\alpha_\mathrm{max}\sim \st^{1.6-1.7}$ as $b$ increases, so $\alpha_\mathrm{max}$ is not sensitive to $b$.  The overall pattern of growth rates do not change significantly as $b$ changes, thus non-linear drag laws have limited effect.

\subsection{Effect of gas compressibility}\label{effects_from_eta}
We briefly examine the effect of gas compressibility by varying  $\hat{\eta}\equiv \eta R/\Hgas$. For fixed $\eta$, i.e. the global radial pressure gradient, $\hat{\eta}\propto 1/c_s$. Then larger $\hat{\eta}$ correspond to smaller $c_s$, i.e. higher compressibility; and vice versa. Thus $\hat{\eta}$ is a measure of gas compressibility \citep[see][and Appendix \ref{analytic_model}]{youdin07b}.  

\begin{figure*}
\begin{center}
\includegraphics[width=\linewidth]{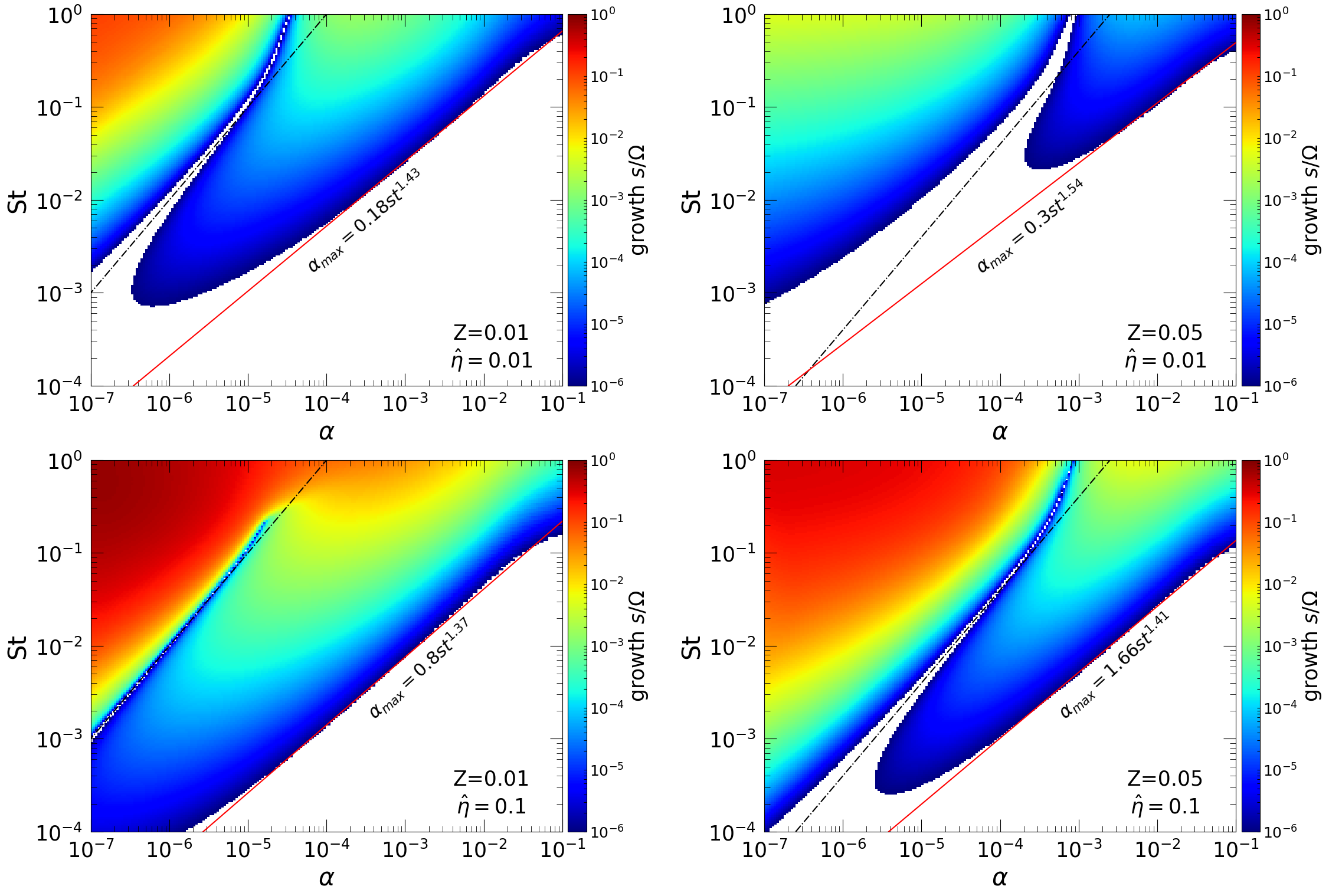}
\end{center}
\caption{Growth rates of the streaming instability at fixed metallicity for different levels of gas compressility, as measured by $\hat{\eta}=0.01$ (upper panels) and $\hat{\eta}=0.1$(lower panels). The black dot-dashed lines in each panels correspond to Eq. Red lines empirical fits to the maximum allowed $\alpha$ for growth rates to remain $>10^{-6}\Omega$.  \ref{alpha_st_relation_dg1}.
} 
\label{fig:eta_z}
\end{figure*}

In inviscid disks, the linear SI is unaffected by gas compressibility \citep{youdin05a}. However, for viscous disks, we find SI growth rates increase with gas compressibility. This is shown in Fig. \ref{fig:eta_z} where growth rates at fixed metallicities are computed for $\hat{\eta}=0.01$ and $0.1$ (recall our nominal value is $0.05$). The comparison between upper and lower panels indicates that larger compressibility leads to higher growth rates for the same $\st$ and $\alpha$. For instance, in the case with $Z=0.01$, $\alpha=10^{-5}$, and $\st=10^{-3}$, the growth rate is less than $10^{-6}\Omega$ with $\hat{\eta}=10^{-2}$ while the growth rate is about $10^{-4}\Omega$ with $\hat{\eta}=10^{-1}$. This is because diffusion appears as the quantity $\mathcal{D}\equiv \delta/\hat{\eta}^2$ in the dimensionless equations (see Appendix \ref{analytic_model}). Thus at fixed $\delta$ increasing $\hat{\eta}$ diminishes the stabilization effect of diffusion.   

Similar to the cases in $\hat{\eta}=0.05$, the differences between left and right panels of Fig. \ref{fig:eta_z} suggest that larger metallicity expand the dust-rich SI (regions above the black dash lines, which represent unit $\epsilon$). Although increasing gas compressibility leads to faster growth, we find the maximum allowed $\alpha$ still approximately scales as $\st^{1.4-1.5}$ (red lines). 

It is important to remember our results apply only to the \emph{linear} phase of the instability. In non-linear regime, \cite{bai10c} in fact find that increasing $\eta$ results in weaker particle clumping. This suggests that in reality there is an optimum $\eta$ that is a balance between linear growth and non-linear clumping that maximizes planetesimal formation. 


%% file: application.tex
\section{Application to protoplanetary disks}\label{application}

We now apply the above linear theory to global models of PPDs. We first extract the  dimensionless input parameters from physical disk models. To do so, we couple classic viscous accretion disk theory, e.g. \cite{pringle81} with typical disk profiles used in the literature. We then compute growth timescales and optimum lengthscales at each radius in order to assess the role of SI in realistic PPDs. 
Unless otherwise stated, we assume a central star of mass $1M_\sun$ with a Keplerian rotation $\Omega\propto R^{-3/2}$ profile.  


We consider the minimum mass Solar nebulae (MMSN) as described in \cite{chiang10}. These disk models have gas surface density and midplane temperature profiles 
\begin{align}
    \sigg &= 2200 F \left(\frac{R}{\mathrm{AU}}\right)^{-3/2} \mathrm{g}\,\mathrm{cm}^{-2}, \\
    T &= 120 \left(\frac{R}{\mathrm{AU}}\right)^{-3/7}\mathrm{K},
\end{align}
where $F$ is a scale factor. Assuming vertical hydrostatic equilibrium and a vertically isothermal equation of state, the midplane gas density and pressure scale height profiles are:  
\begin{align}
    \rhog &= 2.7\times10^{-9}F\left(\frac{R}{\mathrm{AU}}\right)^{-39/14}
    \mathrm{g}\,\mathrm{cm}^{-3},\\
    \Hgas &= 0.022 R \left(\frac{R}{\mathrm{AU}}\right)^{2/7}. 
\end{align} 
We assume a constant metallicity, so the dust surface density is given by  $\Sigma_\mathrm{d}(R) = Z \sigg(R)$.  

Using the above model we can compute global profiles of $\st(R)$, $\hat{\eta}(R)$, $\epsilon(R)$, and $\nu(R)$, as required for linear theory at each radius. We consider a single dust population of a given size and internal density in the Epstein regime. This gives 
\begin{align}
\st = 1.1 \times 10 ^ { - 3 } F^ { - 1 } \left( \frac { R } { \mathrm { AU } } \right) ^ {3 /2 }
\left(  \frac{\rho_\bullet}{\mathrm{g}\mathrm{cm}^{-3}} \right)\left(\frac{a_p}{\mathrm{cm}}\right). 
\end{align}
Similarly, the dimensionless pressure gradient $\hat{\eta}\equiv \eta R\Omega/c_s$ is  
\begin{align}
  \hat{\eta} = 0.035 \left( \frac { R } { \mathrm { AU } } \right) ^ { 2 / 7 }.
\end{align}
Thus $\hat{\eta}$ is almost a constant.  

{ We also compare the growth timescale with radial drift timescale of dust particles. The drift timescale is calculated from the radial drift veloctiy in Eq. \ref{Wx} as 
\begin{align}
\frac{t_\mathrm{drift}}{\mathrm{yr}} =  \frac{0.73}{F} \left(\frac { R } { \mathrm { AU } }\right)^{17/7}+5.8\times 10^{5} F (1+\epsilon)^2 \left(\frac { R } { \mathrm { AU } }\right)^{-4/7},  
\end{align}
where the midplane dust-to-gas ratio $\epsilon$ is calculated according to Eq.  \ref{hd_hg}--\ref{dg_fixZ}, which requires a prescription for the disk viscosity (see below). Notice $t_\mathrm{drift}$ diverges at both small and large radii, as for fixed particle sizes these correspond to $\st\to 0$ and $\st\to \infty$, respectively, which have the slowest drift speeds.  
}

We consider two models of the disk viscosity, $\nu = \alpha c_s\Hgas $, described below, which is then used to compute $\Hdust/\Hgas$ and $\epsilon(R; Z, \st, \alpha)$ from Eq. \ref{hd_hg} and \ref{dg_fixZ}, respectively.


\subsection{Accreting disks}
In this case, we assume turbulence leads to (gas) accretion onto the central star, and the viscosity parameter is a measure of turbulent angular momentum transport. 
Far from the inner disk boundary we have 
\begin{align}
\nu \sigg \simeq \frac{\dot{M}}{3\pi}, 
\end{align}
where the (gas) mass accretion $\dot{M}$ is a constant input parameter. For the above gas disk profiles we obtain    
\begin{align}
  \alpha(R) = 8.8 \times 10 ^ { - 3 } F ^ { - 1 }\left( \frac {R} { \mathrm { AU } } \right) ^ {3/7} \left(\frac{\dot{M}}{10^{-8}M_\sun\mathrm{yr}^{-1}}\right). \label{alpha_accreting}
\end{align}
Thus, higher accretion rates require larger $\alpha$; while higher disk masses imply smaller $\alpha$. Note that the viscous accretion flow has a characteristic radial velocity $\sim \nu/R $, but this is neglected in our local linear analysis.

\begin{figure*}
 \begin{center}
 \includegraphics[width=\linewidth]{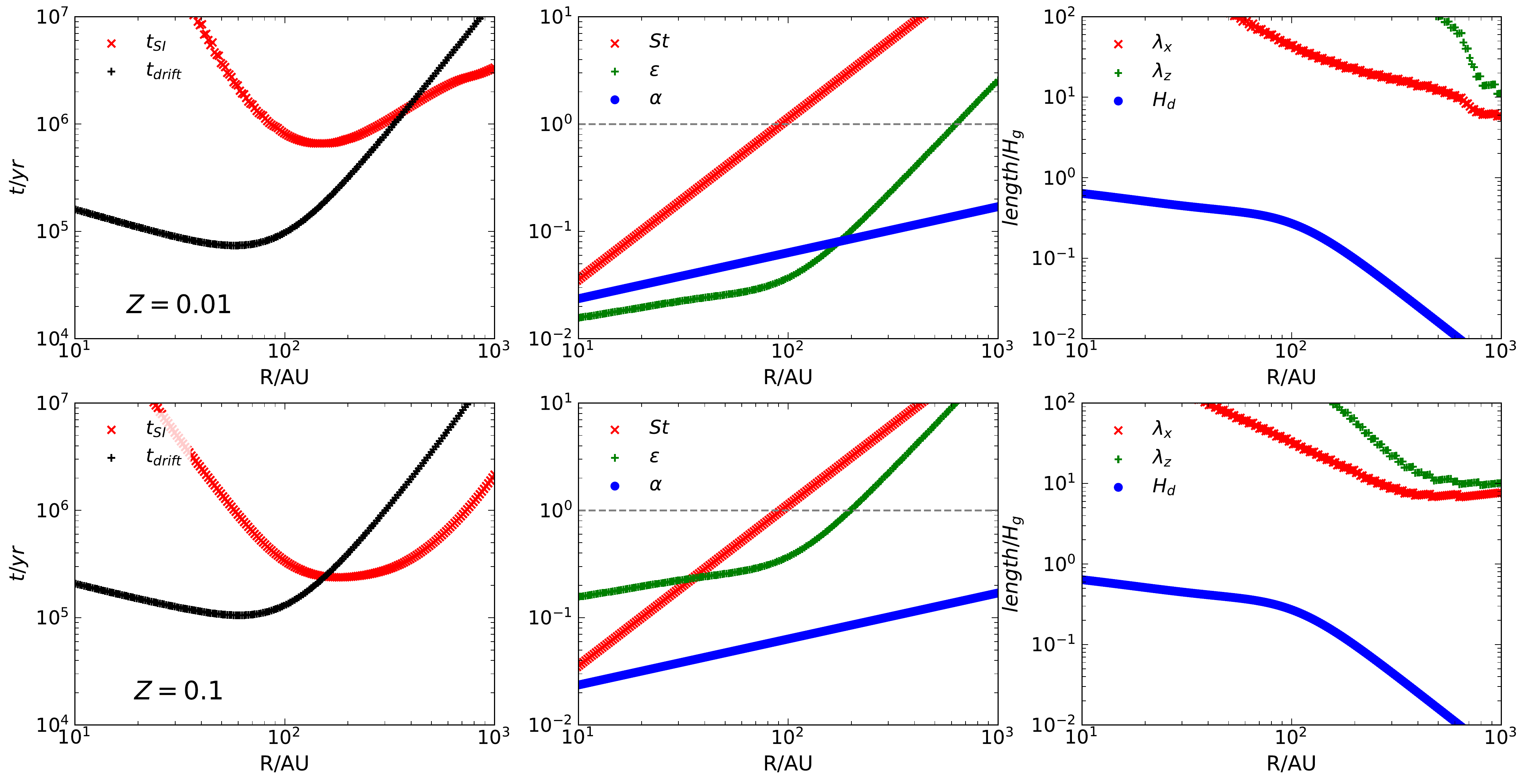}
 \end{center}
 \caption{Streaming instability as a function of radius in a  minimum mass solar nebula, accreting PPD. 
 The left panel shows growth { $\tgrow$ (red) and radial drift timescales $t_\mathrm{drift}$(black)}; the middle panel shows the Stokes number $\st$ (red), midplane dust-to-gas ratio $\epsilon$ (green) and viscosity $\alpha$ (blue); and the right panel shows the radial and vertical wavelengths $\lambda_{x,z}$ (red, green) of the SI mode, in comparison with the dust scale-height $\Hdust$ (blue). The horizontal dashed lines indicate $\st = 1$ and $\epsilon =1$. 
 \label{fig:si_acc}
 }
 \end{figure*}

In Fig. \ref{fig:si_acc} we present results for a standard mass accretion rate $\dot{M} = 10^{-8}M_\sun\mathrm{yr}^{-1}$, disk mass $F=1$, and cm-sized particles with internal density  $1\,\mathrm{g}\,\mathrm{cm}^{-3}$. We consider a solar-metallicity disk with $Z=0.01$ (top) and a dust-enriched disk with $Z=0.1$ (bottom). We plot growth { and dust radial drift timescales (left)}; Stokes number, mid-plane dust-to-gas ratios, $\alpha$-viscosity values (middle); and optimum wavelengths (right). 


\shipout\box255

For both metallicities growth within the disk lifetime ($\lesssim 10^7\mathrm{yrs}$) is only possible for $R\gtrsim 20$--$30\mathrm{AU}$. For $Z=0.01$, growth timescales $\gtrsim 1\mathrm{Myr}$ at all radii. {        Increasing} $Z$ { to $0.1$} allows the SI to grow in $\lesssim 1\,\mathrm{Myrs}$ at a few tens of AU. { However, for $Z=0.1$ we find $\epsilon < 1$ for $R\lesssim 200\mathrm{AU}$}. This is due to the turbulent stirring by gas, with rather large viscosity values of $\alpha\sim O(10^{-2})$ compared to recent theoretical models \citep[e.g.][]{bai14,simon18} and measurements of disk turbulence  \citep{flaherty17,flaherty18}. It is doubtful that such dust-poor conditions can lead to planetesimal formation in the nonlinear regime of the SI \citep{johansen09}. { Furthermore, radial drift timescales in these regions are shorter than the growth time, implying dust may be lost to the star before significant growth.}

{ On the other hand, we find $\tgrow \lesssim t_\mathrm{drift}$ for $R \gtrsim 400\mathrm{AU}$ for $Z=0.01$ and for $R \gtrsim 100\mathrm{AU}$ for $Z=0.1$, suggesting that dust in the outer disk can undergo efficient SI before falling into the host star. There is, however, some uncertainty because in these regions $\st\gtrsim 1$, which violates the fluid treatment of dust.}

We find characteristic SI lengthscales are of $O(10\Hgas)$ in both cases. This is problematic in two respects. At $\sim 100\mathrm{AU}$ the gas disk aspect-ratio is $\Hgas/R \sim 0.08$, implying a radial length scale 
comparable to the disk radius, $\lambda_x \sim 8\Hgas \sim R $. At this scale, the global disk geometry may become important, but this is neglected in our local stability analyses. Moreover, we find vertical lengthscales are much larger than the dust  scale-height, $\lambda_z\gg \Hgas\gg \Hdust$. The existence of such vertically-extended modes may then depend on physical conditions at the surface of the dust layers. This issue is beyond the scope of this work.

\subsection{Non-accreting disks}
These { disk} models may be { considered as} representing `dead zones' in PPDs, where  weak turbulence result from hydrodynamic instabilities, but do not contribute to mass accretion \citep{bai16}. 
{ Thus we require $\nu\neq 0$, but without a corresponding radial gas flow ($v_R=0$). This is in fact consistent with our local models.} 
 { Since we have chosen MMSN surface density profiles, we deduce the appropriate viscosity profile as follows. It turns out these disks have almost a constant $\alpha$ that may be specified independently.  
}

We recall from classic viscous theory that the (gas) radial velocity is given via 
\begin{align}
 R \sigg v_R \frac{d}{dR}\left(R^2\Omega\right) = \frac{d}{dR}\left(R^3\nu\sigg\frac{d\Omega}{dR}\right).
\end{align} 
(Note that this equation is also applicable to unsteady disks.) For a Keplerian disk we obtain 
\begin{align}    
        v_R =-\frac{3}{\sqrt{R}\sigg}\frac{d}{dR}\left(\nu \sigg \sqrt{R}\right).
\end{align}
Thus, if $\nu\sigg\sqrt{R}$ is constant then $v_R=0$. { Such non-accreting disks have also been employed in other problems, e.g. disk-planet interaction \citep{paardekooper09d}.}
We use { the above} constraint to set  
\begin{align}
  \alpha(R)  = \alpha_1 \left( \frac { R } { \mathrm { AU } } \right) ^ {-1 / 14 },
\end{align}
where $\alpha_1$ is the viscosity coefficient at $1\mathrm{AU}$, and is an input parameter. 

\begin{figure*}
 \begin{center}
 \includegraphics[width=\linewidth]{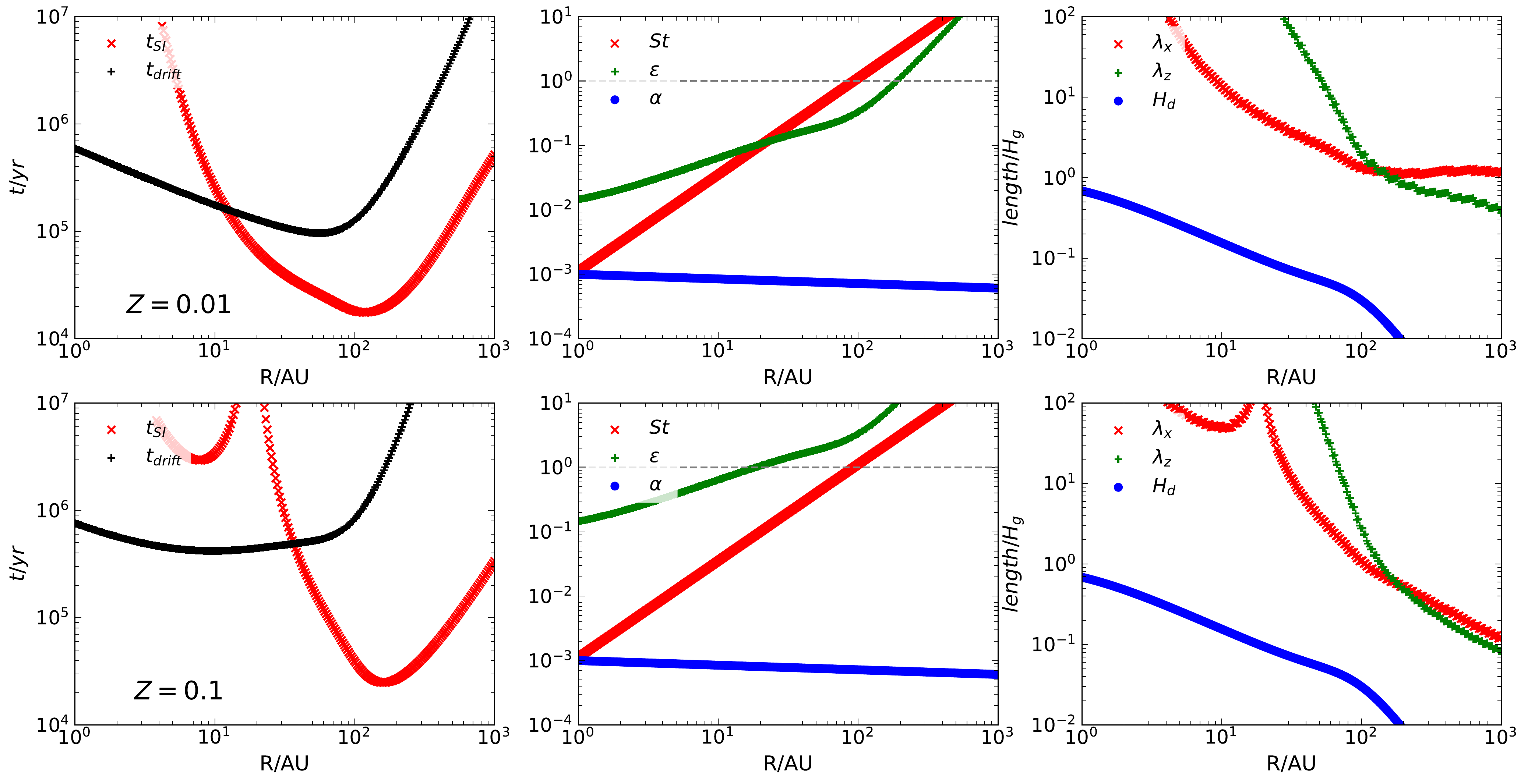}
 \end{center}
 \caption{Same as Fig. \ref{fig:si_acc} but for non-accreting disks parameterized by $\alpha_1=10^{-3}$. 
 \label{fig:si_nonacc}
 }
 \end{figure*}

Fig. \ref{fig:si_nonacc} show example results in non-accreting disks. Here, we fix $\alpha_1=10^{-3}$. We find that SI can grow within the disk lifetime for $R\gtrsim 3\mathrm{AU}$, much smaller than in the accreting disks. This is due to the smaller viscosity compared to the non-accreting disk above. However, as in accreting disks, the SI grows fastest at $R \sim 100\mathrm{AU}$. 




For $Z=0.01$ we find $\epsilon \lesssim 1$ for $R\lesssim 100\mathrm{AU}$.  
Here, the linear SI grows sufficiently fast, but $\epsilon$ is insufficient for dust clumping \citep{johansen09}. Thus we only expect the development of axisymmetric dust rings. 

On the other hand, in the $Z=0.1$ disk we find $\epsilon\gtrsim 1$ for $R\gtrsim 20\mathrm{AU}$, and growth timescales are much shorter than the disk lifetime. In this case the SI will likely lead to planetesimal formation. The sudden increase of $\tgrow$ at $R\sim20$AU is associated with $\epsilon$ approaching unity, whence SI is quenched (see, e.g. Appendix \ref{analytic_dust_rich}). 

{ In both cases, a comparison between radial drift timescales (black) and SI growth timescales (red) are also shown. We find $\tgrow\lesssim t_\mathrm{drift}$ beyond $10\mathrm{AU}$ and $30\mathrm{AU}$ for $Z=0.01$ and $Z=0.1$, respectively. Here the SI can grow before dust particles are lost to the star. Conversely, interior to these radii SI is limited by radial drift instead of disk lifetimes.}

In either disk models we find $\lambda_x\sim \Hgas\ll R$ at $R=100\mathrm{AU}$, which is consistent with local analysis. However, vertical wavelengths are still significantly larger than $\Hdust$. { We have attempted to restrict $\lambda_z \leq 2\Hdust$ (the full dust layer thickness) for self-consistency}, but this in fact lead to growth timescales exceeding the disk lifetime. This is because viscosity is effective in stabilizing perturbations with such small vertical lengthscales. 


\subsection{Dust rings around HL Tau}

We now apply the above disk models to examine whether or not the SI can explain the formation of the dust rings and gaps observed in the PPD around HL Tau \citep{alma15}. These dust rings are located between $\sim 13$ to $\sim 91$AU, and adjacent rings are separated by $\sim 10$ --- $20$AU. 
We require growth timescales $\lesssim 1\mathrm{Myr}$ for consistency with HL Tau's young age. Since the disk mass of HL Tau is about $0.1M _ { \odot }$ \citep{alma15}, which is about 10 times larger than MMSN, we adopt $F=10$ in the calculations below. We still consider cm-sized particles with internal density $1\,\mathrm{g}\,\mathrm{cm}^{-3}$.  Calculations with mm-sized particles yield essentially no growth within the disk's age. 

For the accreting disk model we adopt $\dot{M}~\sim~10^{-7}M_\odot\mathrm{yr}^{-1}$ \citep{beck10}. However, this implies a high viscosity, with $\alpha$ increasing from $\sim 10^{-2}$ at 1AU to $\sim 0.07$ at 100AU. As a result, SI growth timescales exceed $100\mathrm{Myr}$. We thus discard this model. 

\begin{figure*}
 \begin{center}
 \includegraphics[width=\linewidth]{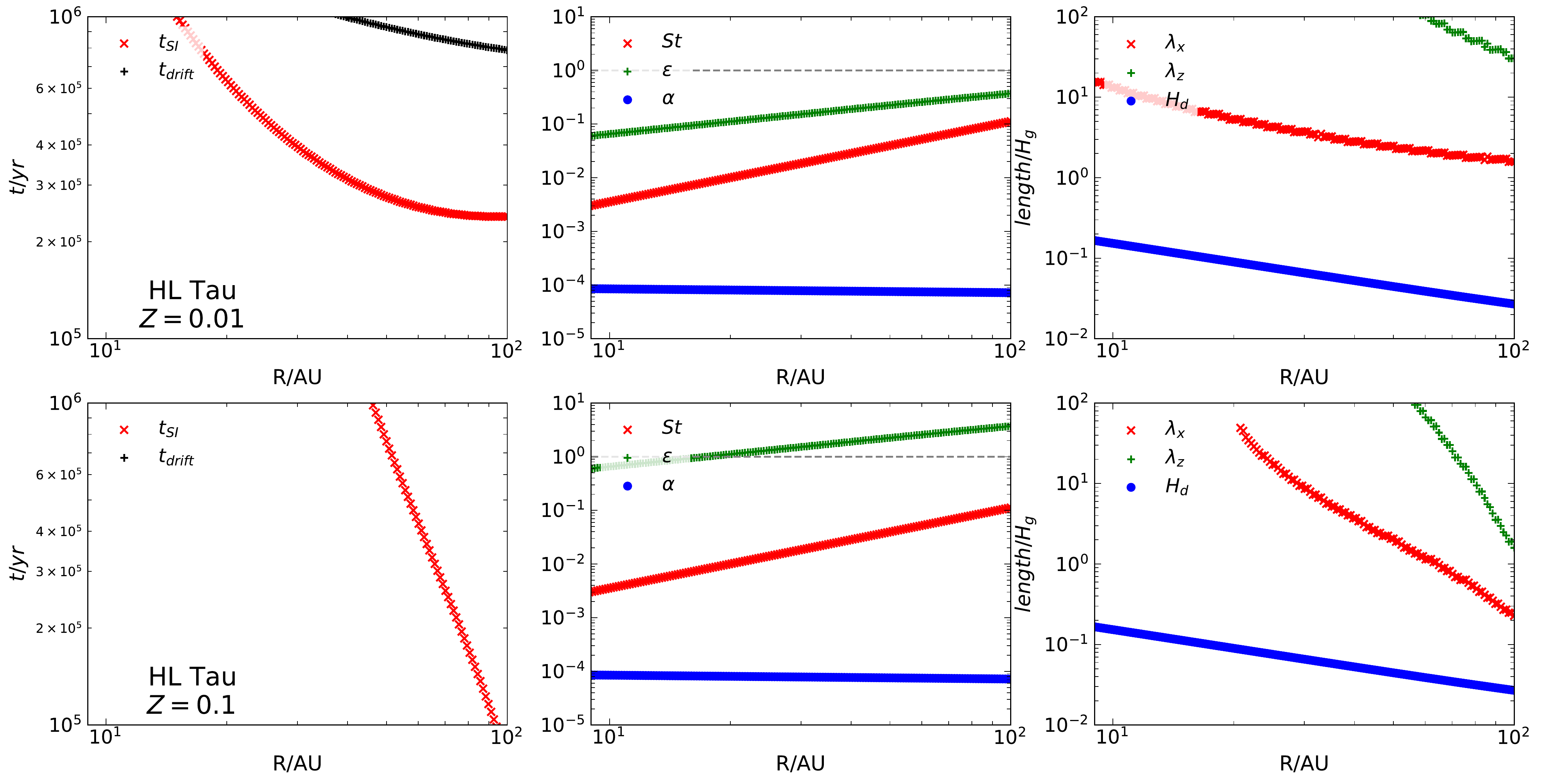}
 \end{center}
 \caption{Streaming instability in the PPD around HL Tau, based on the non-accreting disk model. 
 \label{fig:si_hltau}}
 \end{figure*}

For the non-accreting disk model, we set $\alpha_1 = 10^{-4}$. This is motivated by the estimate made by \cite{pinte16} based on observational constraints on the dust layer thickness in the HL Tau disk. Results are shown for $Z=0.01$ and $0.1$ in Fig. \ref{fig:si_hltau}. For both cases the fastest growth occurs near the outer disk edge $\sim 100$AU. { Notice here that radial drift does not limit the instability growth. (For $Z=0.1$, $t_\mathrm{drift}$ exceeds a million years, so it is outside the plotted range.)}
  
For $Z=0.01$, we find $\tgrow\lesssim 1\mathrm{Myr}$ beyond $\sim 15$AU. In these regions,  The MMSN profile gives $\Hgas \simeq 0.7$AU at $R=15$AU. Here, the radial wavelength is $10\Hgas = 7 $AU, which is too small compared to the observed ring separations. By contrast, at $100$AU the radial wavelength is $\simeq 2\Hgas = 16$AU, which is broadly consistent with observations. In the $Z=0.1$ disk we find SI only grows sufficiently fast for $R \gtrsim 50$AU. However, at these radii the radial wavelengths range between $6.7$AU at $50$AU to $1.6$AU at $100$AU, which are too small compared to observations. 

From the above crude comparison, we conclude that the $Z=0.01$ disk can plausibly explain dust rings observed in the outer disk ($R\gtrsim 70$AU where $\lambda_x \gtrsim 10$AU). However, we again find $\lambda_z \gg \Hdust$. { Additional calculations enforcing $\lambda_z \leq  2\Hdust$ resulted in decaying modes, which would play no role.}    

{ In any case}, because $\epsilon\lesssim 1$ { in the $Z=0.01$} disk model, { SI is unlikely to result in planetesimal formation}. This would suggest that planetesimals cannot form via the SI in the HL Tau disk, leaving only axisymmetric rings from the { instability}.  

%% file: summary.tex
\section{Summary and discussion}\label{summary}

In this paper we assess the efficiency of planetesimal formation via the streaming instability (SI) in physical models of protoplanetary disks (PPDs). To this end, we generalize the linear theory of the SI to include disk turbulence, modelled as a gas viscosity, with a corresponding particle stirring modelled by dust diffusion. We also explore the modest effect of non-linear drag laws and gas compressibility on the SI. For the most part we adopt the standard two-fluid model of dusty gas, but also verify some calculations with a simplified, one-fluid model generalized from \cite{lin17} to include dust diffusion. 


We find the SI is sensitive to turbulence. Gas viscosity and particle diffusion stabilizes the SI and increases its characteristic lengthscale, as expected on physical grounds. SI with small particles are effectively stabilized by turbulence. For example, at fixed dust-to-gas ratios for $\st\sim 10^{-2}$ growth rates become negligible for $\alpha\gtrsim 10^{-3}$. We also find the SI is more easily smeared out in the vertical direction than in the radial direction, consistent with \cite{umurhan19}.

In a physical disk, however, turbulence also changes the background disk structure, namely the equilibrium dust-to-gas ratio. Accounting for this yield two regimes: at low viscosity the dust-to-gas ratio exceeds unity and the SI grows on dynamical timescales; and at high viscosity the dust-to-gas ratio falls below unity and the SI grows slowly, eventually exceeding timescales of interest. Our numerical results indicate a convenient scaling for the maximum viscosity as $\alpha_\mathrm{max}\sim \st^{1.5}$. 

We apply linear stability analysis to global models of PPDs. We consider the standard minimum-mass Solar nebula disk models with viscosity either chosen to yield a specified global gas accretion rate, or set independently in a non-accreting disk. Even considering large, cm-sized particles that should favor the SI, we find the SI only grows within typical disk lifetimes of a few Myrs outside $\sim 10$AU. The SI is most efficient around $100$AU, where growth timescales can approach $O(10^4)$yrs at low viscosity. 

On the other hand, we consistently find vertical lengthscales of the SI exceeds the dust layer thickness. { Taken at face value, this suggests that in viscous disks the SI has little vertical structure across the dust layer. However, only a stratified linear stability analysis can confirm whether not there exists vertically unstructured modes, or modes that can be confined to the dust layer} (Lin et al., in preparation).  


Our local analyses also neglect the viscosity-induced gas accretion flow that exist in global disks. Future work should account for this by calculating the background equilibrium flow self-consistently. This gas accretion flow would drive an additional relative drift between dust and gas, on top of that due to dust-gas drag. As the relative dust-gas drift is the culprit of dust-gas instabilities, including the SI  \citep{squire17}, we can expect a gas accretion flow to also affect the SI. We leave this to future work. 







%% file: appendix.tex
\section{One-fluid model for dusty gas with particle diffusion}\label{one_fluid_model}
Here we describe the `one-fluid' model of dusty gas developed by \cite{lin17}, based on the earlier studies of \cite{laibe14} and \cite{price15}. We extend these models by including dust diffusion, but neglect gas viscosity for simplicity \citep[see][for a viscous, but diffusionless version of the one-fluid model]{lovascio19}. 

In this description one forgoes separate gas and dusty variables and work with the total density
\begin{align}
  \rho \equiv \rhod + \rhog, 
\end{align}
and the center of mass velocity of the mixture
\begin{align}
\bm{U} \equiv \frac{\rhod\bm{W} + \rhog\bm{V}}{\rho}. 
\end{align}
Furthermore, by considering small particles one applies the `terminal velocity approximation' \citep{youdin05a,jacquet11} so that
\begin{align}
  \bm{W} - \bm{V} = \frac{\nabla P}{\rhog}\tstop, \label{term_vel_approx}
\end{align}
where $\tstop \equiv \taus\fg$ is the relative stopping time and $\fg=\rhog/\rhog$ is the gas fraction. (It is simpler to work with $\tstop$ in the one-fluid framework.) As before we consider strictly isothermal gas so that $P = c_s^2\rhog = 
c_s^2(1-\fdust)\rho$, where $\fdust = \rhod/\rho$ is the dust-fraction. 

With the above definitions and approximations, the two-fluid equations (\ref{gas_mass}---\ref{dust_mom}) can be combined and simplified to give 
\begin{align}
 &\frac{\p \rho}{\p t} + \nabla\cdot\left(\rho\bm{U}\right)=\nabla\cdot\left[D P \nabla\left(\frac{\rho}{P}\right)\right] , \label{mass0}\\ 
  &\frac{\p\bm{U}}{\p t} + \bm{U}\cdot\nabla\bm{U} = -
  \frac{1}{\rho}\nabla  P - \OmK^2 R \hat{\bm{R}}\,\label{mom0}\\ 
 & \frac{\p P}{\p t} + \bm{U}\cdot\nabla P 
  = - P \nabla\cdot\bm{U}  + c_s^2\nabla\cdot\left(\tstop \fdust\nabla P\right). \label{energy0} 
\end{align} 
Note that only three of the four evolutionary equations remain because the dust velocity has been eliminated with the terminal velocity approximation. Eq. \ref{energy0} is an effective energy equation that result from the dust continuity equation (\ref{dust_mass}). Terms of $O(\tstop^2)$ are neglected due to the assumption of small particles. For the following derivations, it is convenient to define  
\begin{align}
  \mathcal{C}&\equiv c_s^2\nabla\cdot\left(\tstop \fdust\nabla
    P\right) = -c_s^2\nabla\cdot\left(\rho\tstop\fdust\bm{F}\right),\\
  \bm{F} &\equiv - \frac{\nabla P}{\rho}. 
\end{align}

\subsection{Relative stopping time}
We adopt the same stopping time parameterization as the two-fluid model (Eq. \ref{taus_def}). The corresponding definition of the relative stopping time is 
\begin{align}
  \tstop 
  = t_\mathrm{s,eqm}\frac{\left[\rho\rho_\mathrm{g}^{a-1}\left|\bm{W}-\bm{V}\right|^b\right]_\mathrm{eqm}}{\rho\rhog^{a-1}\left|\bm{W}-\bm{V}\right|^b}. \label{tstop_def}  
\end{align} 
Combining Eq. \ref{term_vel_approx} with  Eq. \ref{tstop_def} gives 
\begin{align}
|\bm{W} - \bm{V}|^{1+b} = t_\mathrm{s,eqm} 
\left[ \rho_\mathrm{g}^{a-1}\rho|\bm{W}-\bm{V}|^{b}\right]_\mathrm{eqm}
\frac{1}{\rhog^a}\left|\frac{\nabla P}{\rho}\right|\label{wv_expression}.
\end{align}
It will prove useful to have an expression for $\rho\tstop$. Using Eq. \ref{wv_expression}, we find 
\begin{align}
  \ln{\left(\rho\tstop\right)} =
  \left(1 - \frac{a}{1+b}\right)\ln{\rhog} -
  \frac{b}{1+b}\ln{\left|\frac{\nabla P }{\rho}\right|} + \mathrm{const}. \label{tstop_1f}
\end{align}
\subsection{One-fluid equilibrium}
The one-fluid momentum equations admit axisymmetric, steady state equilibrium solutions with $U_R = U_Z=0 $, and 
\begin{align}
  U_\phi^2 &= R^2\OmK^2(R) + \frac{R}{\rho}\frac{\p P}{\p R},\\
  & = R^2\OmK^2(1 - 2\fg\eta).   
\end{align} 
Since $\eta\ll 1$ we can set $U_\phi= R\OmK$ in practice.   

The steady state continuity and effective energy equations (\ref{mass0}, \ref{energy0}) are generally not satisfied for arbitrary density and pressure profiles. However, if we consider power law disks where the density and pressure varies on a global scale $\p_R\sim 1/R$, then the effective energy equation implies the background would evolves on a timescale of $O(1/h^2\st \OmK)$, where $h\ll 1$ is the disc aspect-ratio. Similarly, the continuity equation gives a background evolution timescale of $O(1/\delta h^2 \OmK)$. These timescales are much longer than the SI growth timescales found below. We can thus self-consistently neglect the background evolution.   


\subsection{Linearized one-fluid equations}
As in the main text we consider axisymmetric Eulerian perturbations with space-time dependence
$\exp{\left(\sigma t + k_xR + k_zz\right)}$ and linearize Eq. \ref{mass0}---\ref{energy0}. We assume the radial wavenumbers $|k_xR|\gg 1$ so that background density and pressure gradients may be neglected  when compared to that in the perturbed variables. We find 
\begin{align}
  \sigma \frac{\dd\rho}{\rho} & = -\ii k_x \dd U_R - \ii k_z \dd U_z 
     - D
    k^2\left[\frac{\dd\rho}{\rho} - (1+\epsilon)\frac{\dd
        P}{c_s^2\rho}\right], \label{lin_mass}\\  
  \sigma\dd U_R  & = 2\Omega\dd U_\phi + \dd F_R,\\
  \sigma\dd U_\phi &=-\frac{\Omega}{2}\dd U_R,\\
  \sigma\dd U_z &= -\ii k_z \frac{\dd P}{\rho}\\
 \sigma\frac{\dd P}{\rho} &= -\frac{P}{\rho}\left(\ii k_x \dd U_R + \ii k_z
               \dd U_z\right)
               + \frac{\dd\mathcal{C}}{\rho},\label{lin_energy}
\end{align}
where $k^2 = k_x^2 + k_z^2$ and 
\begin{align}
  \dd F_R = - F_R\frac{\dd\rho}{\rho} - \ii k_x \frac{\dd P}{\rho}.
\end{align}
The linearized diffusive term is 
\begin{align}
  -\frac{\dd\mathcal{C}}{c_s^2\rho} = & \ii k_x F_R \tstop
  \left[\fg\frac{\dd\rho}{\rho} - \frac{1}{c_s^2}\frac{\dd P}{\rho} +
    \fdust\frac{\dd\left(\rho\tstop\right)}{\rho\tstop}\right]
  + \tstop\fdust\left[\left(k_x^2+k_z^2\right)\frac{\dd P}{\rho} -
    \ii k_x F_R \frac{\dd\rho}{\rho}\right], 
\end{align}
and linearizing Eq. \ref{tstop_1f} gives
\begin{align}
  \frac{\dd\left(\rho\tstop\right)}{\rho\tstop}=  \frac{1}{\fg c_s^2}\left(1 -
    \frac{a}{1+b}\right)\frac{\dd P}{\rho}  -
  \frac{b}{1+b}\frac{\dd F_R}{F_R}. 
\end{align}
The full expression for $\dd\mathcal{C}$ is then
\begin{align*}
-\frac{\dd\mathcal{C}}{\rho c_s^2} &=  \ii k_x F_R \tstop \left(\fg -
  \frac{\fdust}{1+b}\right)\frac{\dd\rho}{\rho}
+ \tstop \left[\frac{\ii k_x F_R}{c_s^2}\theta + 
\fdust\left(k^2 - \frac{b k_x^2}{1+b}\right)
\right]\frac{\dd P}{\rho},
\end{align*}
where 
\begin{align}
\theta \equiv \epsilon\left(1 -
  \frac{a}{1+b}\right) - 1.
\end{align}

\begin{figure}
\begin{center}
\includegraphics[width=\linewidth]{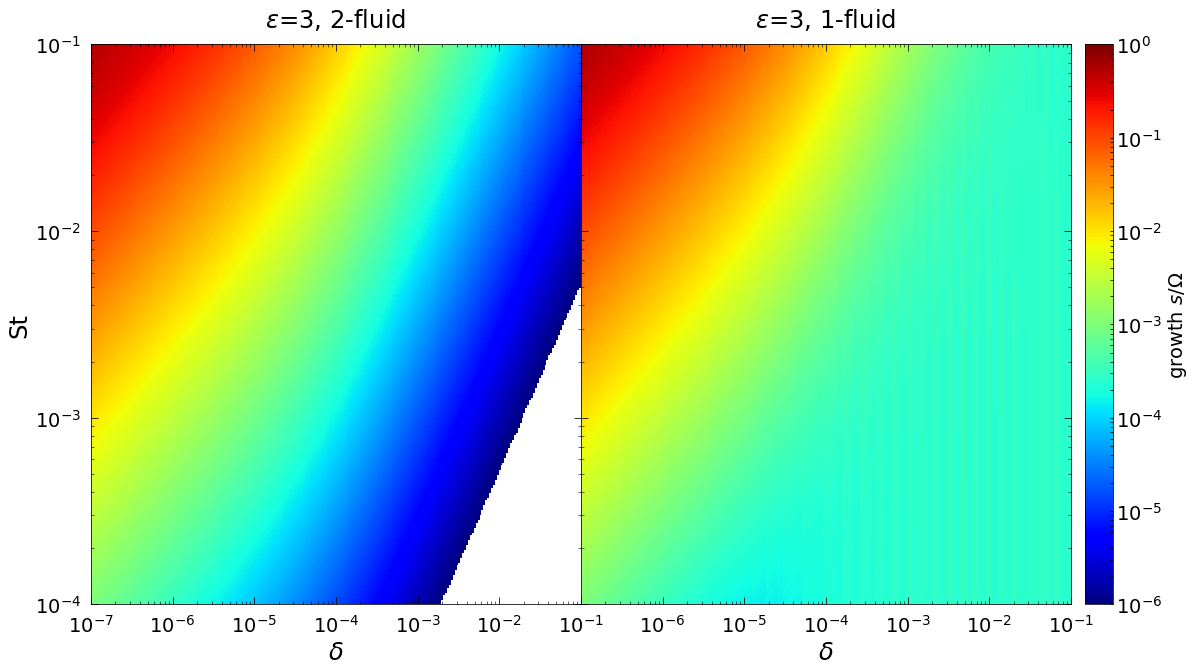}
\end{center}
\caption{Growth rates as a function of Stokes number $\st$ and dust diffusion $\delta$ computed from the two-fluid (left) and one-fluid (right) models. The optimal $K_x$ and $K_z$ values found in the two-fluid model are used in the one-fluid model. Plots are truncated if growth rates are smaller than $10^{-6}\Omega$.
\label{fig:B2}}
\end{figure}

In Fig. \ref{fig:B2} we compare growth rates from the full two-fluid equations and the one-fluid model. We consider Epstein drag with $a=1, b=0$. For the two-fluid model, we include particle diffusion but neglect gas viscosity. We also use the optimum wavenumbers found in the two-fluid model in the one-fluid calculation. We find 
the one-fluid model is consistent with the two-fluid model when diffusion is weak with $\delta\lesssim 10^{-4}$. The one-fluid framework produces spuriously growing modes for larger $\delta$; although their growth rates remain small. 

\subsection{Reduced model}\label{analytic_model}

We can simplify the linearized one-fluid equations to obtain a reduced dispersion relation. It is convenient to first render the one-fluid equations dimensionless by adopting dimensionless parameters $ 
\st = \taus\Omega, \,\, K_{x,z} = \eta R k_{x,z}$ as used in the main text. (Recall the Stokes number $\st$ is defined via the particle stopping time $\taus = \tstop/\fg$.) We further introduce the dimensionless variables 
\begin{align}
	n \equiv \ii \sigma/\Omega, \,\, \mathcal{D} \equiv \frac{D}{\left(\eta R\right)^2\Omega}, 
\end{align} 
Note that the dimensionless diffusion coefficient used in the main text is given by $\delta \equiv D/c_s\Hgas = \mathcal{D}\hat{\eta}^2$. 

We eliminate the velocity perturbations from Eq. \ref{lin_mass}---\ref{lin_energy} to obtain 
\begin{align}
&\left\{\zeta_1\left(\ii n - \mathcal{D}K^2\right) - \st\fg\left[\frac{\fdust}{\hat{\eta}^2}\left(K^2 - \frac{b K_x^2}{1+b}\right)+ 2\zeta_1\ii K_x \fg \theta\right]\right\}\frac{\dd P}{c_s^2\rho} = \fg\left[\left(\ii n - \mathcal{D}K^2\right) + 
2\ii K_x\fg\st\left(\fg - \frac{\fdust}{1+b}\right)
\right]\frac{\dd\rho}{\rho},\label{one_fluid_disp1}\\
&\left\{n^2\left[\left(\zeta_2n^2 - 1\right) + 2\ii K_x \fg\right] + \ii n \left(n^2-1\right)\mathcal{D}K^2\right\}\frac{\dd\rho}{\rho} 
= \left[\frac{1}{\hat{\eta}^2}\left(n^2K^2 - K_z^2\right)+\ii\zeta_1n \left(n^2-1\right)(1+\epsilon)\mathcal{D}K^2\right]\frac{\dd P}{c_s^2\rho},\label{one_fluid_disp2}
\end{align}
where we have used the definition $F_R = 2\fg\eta R \OmK^2$, and recall $\hat{\eta} = \eta R\OmK/c_s$. The artificial coefficients $\zeta_{1,2}$ are nominally unity, but are inserted to make the following simplifications:
\begin{itemize}
\item We consider incompressible gas with $\hat{\eta}\ll1$. This is equivalent to setting $\zeta_1\to 0$. Note that this removes the dependence on the gas density via the drag law. 

\item We consider low frequency modes with $|n|\ll 1$. This is equivalent to setting $\zeta_2\to0$. 
\end{itemize} 
These approximations give the dispersion relation 
\begin{align}
&\left(1 + \fdust \st\mathcal{D}K^2 b^\prime\right)n^3 
+\left[\ii\left(\fdust\st b^\prime +\mathcal{D}K^2\right)
+2K_x\fg\st\left(1 - \frac{\fdust}{1+b} - \frac{b\fdust}{1+b}\frac{K_x^2}{K^2}\right)
\right]n^2 - \left(\fdust\st\mathcal{D}K^2b^\prime + \frac{K_z^2}{K^2}\right)n\notag\\
&+\left[2\fg K_x\st\left(\frac{\fdust}{1+b}-\fg\right) - \ii \mathcal{D}K^2\right]
\frac{K_z^2}{K^2}  = 0,\label{cubic_disp}
\end{align}
where 
\begin{align}
b^\prime \equiv 1  - \frac{b}{1+b}\frac{K_x^2}{K^2}. 
\end{align}
Note that for fixed $\delta$ we also require $\delta \ll \mathrm{min}\left[\st \fdust\fg, \left|n - K_z^2/nK^2\right|/(1+\epsilon)\right]$. { Eq. \ref{cubic_disp} agrees with Eq. 97 of \cite{lin17} in the diffusionless limit with Epstein drag ($\mathcal{D}=b=0$) and $|n|\ll 1$. The latter dispersion relation was also derived by  \cite{jacquet11}, \cite{laibe14}, and is consistent with the original SI analysis of \citetalias{youdin05a}.}

\begin{figure}
\begin{center}
\includegraphics[width=\linewidth]{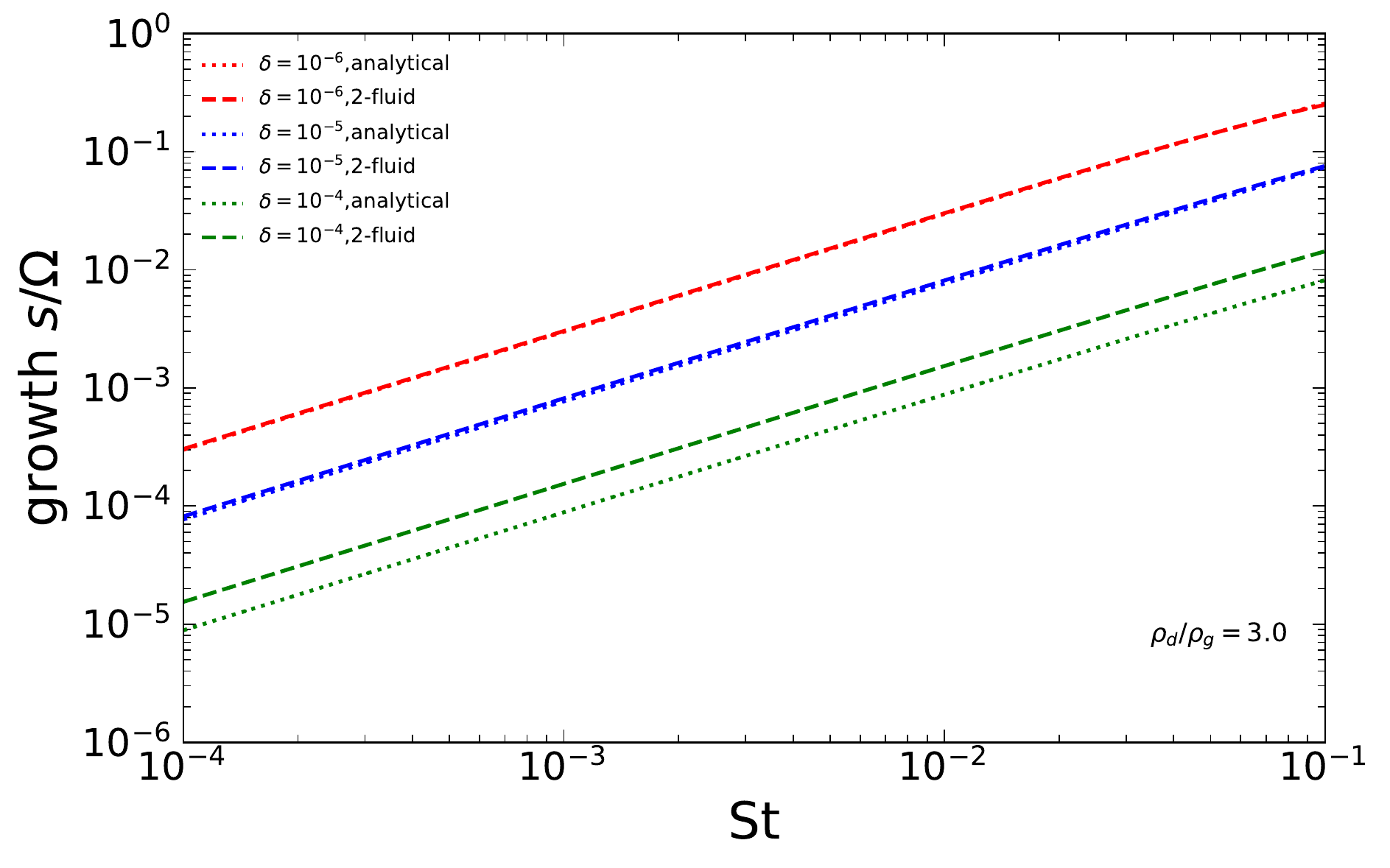}
\end{center}
\caption{Growth rates of analytical model (Eq. \ref{cubic_disp}) and the full two-fluid model with dust-to-gas ratio $\epsilon=3$, as a function of Stokes numbers for different strengths of particle diffusion. We use the optimum values of $K_x$ and $K_z$ found by the two-fluid model in both calculations. 
} \label{fig:B1}
\end{figure}

Fig. \ref{fig:B1}  compares the growth rates from this reduced model (Eq. \ref{cubic_disp}) and the full two-fluid equations, as a function of $\st$ for $\delta=10^{-4},10^{-5}$ and $10^{-6}$. We fix $\rhod/\rhog=3$. We use the optimum $K_x$ and $K_z$ from the two-fluid model when solving Eq. \ref{cubic_disp}. We obtain excellent agreement with the two-fluid model for weak diffusion, $\delta\leq 10^{-5}$. Even for relatively large $\delta=10^{-4}$ the curves are similar. We remark that the incompressible approximation filters out spurious modes in the one-fluid equations that appear at large $\delta$ and small $\st$ (i.e. the lower right part of the right panel in Fig. \ref{fig:B2}). 


\subsection{Dust-rich, diffusionless solutions}\label{analytic_dust_rich}
We can obtain closed-form solutions to Eq. \ref{cubic_disp} in the limit of vanishing dust diffusion. We fix
$K_z, \fdust$ and maximize growth rates over $K_x$, assuming that the optimum 
  $K_x$ is large. (This in turn implies a small stopping time.) Thus 
  we replace $K^2 \to K_x^2$.   

As a simplification we neglect the quadratic term in
Eq. \ref{cubic_disp}. This is consistent with the final solution
obtained below if 
\begin{align}
\fdust \gg \frac{12(b+1)}{12b + 17}. 
\end{align}
Although this can only be marginally satisfied (since $
\fdust < 1$), we find this approximation nevertheless give the correct
growth rates. 

Deleting the quadratic term Eq. \ref{cubic_disp}, we 
solve the depressed cubic 
\begin{align}
n^3 - \mathcal{P}n = \mathcal{Q}, 
\end{align}
{ where the coefficients $\mathcal{P}, \mathcal{Q}$ can be read off Eq. \ref{cubic_disp} with $\mathcal{D}=0$.} 
Writing $n = \mu + \mathcal{P}/3\mu$, we obtain a quadratic for $\mu^3$, 
\begin{align}
\mu^6 - \mu^3 + \frac{\mathcal{P}^3}{27} = 0. 
\end{align}
It turns out the last term may be effectively neglected \citep[][ their Appendix D2]{lin17}, so that $\mu\simeq \mathcal{Q}^{1/3}$. This gives 
\begin{align*}
\real{(n)} 
&= \frac{1}{2}\left(\calA \st^{1/3}K_x^{-1/3} + \calB
    \st^{-1/3}K_x^{-5/3}\right),\\
\imag{(n)}
&= \frac{\sqrt{3}}{2}\left(\calA \st^{1/3}K_x^{-1/3} - \calB
    \st^{-1/3}K_x^{-5/3}\right),\\
 \calA &= \left[2\fg \left(\frac{b+2}{b+1}\fdust
      -1\right)K_z^2\right]^{1/3} ,\\
  \calB&= \frac{K_z^2}{3 A}.
\end{align*}

We maximize growth rates $\imag{(n)}$ over $K_x$. Denoting the associated
quantities with $_*$, we find
\begin{align}
\imag{(n_*)} &=
  \frac{2}{\sqrt{5}}\left(\frac{3}{5}\right)^{3/4}\sqrt{2K_z\st\fg\left(\frac{b+2}{b+1}\fdust-1\right)},\label{dust_rich_grow}\\
\real{(n_*)} &= \frac{\sqrt{3}}{2}\imag{(n_*)},\\
K_{x,*} &= \frac{2}{\sqrt{5}}\frac{K_z}{\imag{(n_*)}}. 
\end{align}
{ The maximum growth rate in Eq. \ref{dust_rich_grow} generalizes Eq. 111 of \citet{lin17} to nonlinear drag laws. Growth} rates vanish when $\fdust = (b+1)/(b+2)$, which is equivalent to $\epsilon = 1+b$. This can also be seen from Eq. \ref{cubic_disp}: in the diffusionless limit when $\epsilon=1+b$ the last term vanishes and the dispersion relation reduces to a quadratic in $n$. From Eq. \ref{dust_rich_grow} we find 
\begin{align}
\frac{\p\imag{(n_*)}}{\p b} = -\frac{\fdust
  \imag{(n_*)}}{2(b+1)\left[(b+2)\fdust - (b+1)\right]}. 
\end{align}
Hence for growing modes with $\imag(n_*)>0$, increasing $b$ tends to stabilize
modes. The effect is small since $\imag{(n_*)}$ only depends on $b$
  through $(b+2)/(b+1)$, which ranges between one and two for
  $b=0$ to $\infty$ (see Eq. \ref{dust_rich_grow}).



%% file: paper.bbl
\begin{thebibliography}{}
\expandafter\ifx\csname natexlab\endcsname\relax\def\natexlab#1{#1}\fi
\providecommand{\url}[1]{\href{#1}{#1}}

\bibitem[{{ALMA Partnership} {et~al.}(2015){ALMA Partnership}, {Brogan},
  {P{\'e}rez}, {Hunter}, {Dent}, {Hales}, {Hills}, {Corder}, {Fomalont}, \&
  {Vlahakis}}]{alma15}
{ALMA Partnership}, {Brogan}, C.~L., {P{\'e}rez}, L.~M., {et~al.} 2015, \apjl,
  808, L3

\bibitem[{{Armitage} {et~al.}(2016){Armitage}, {Eisner}, \&
  {Simon}}]{armitage16}
{Armitage}, P.~J., {Eisner}, J.~A., \& {Simon}, J.~B. 2016, \apjl, 828, L2

\bibitem[{{Auffinger} \& {Laibe}(2018)}]{auffinger17}
{Auffinger}, J., \& {Laibe}, G. 2018, \mnras, 473, 796

\bibitem[{Bai(2014)}]{bai14}
Bai, X.-N. 2014, The Astrophysical Journal, 798, 84

\bibitem[{{Bai}(2015)}]{bai15}
{Bai}, X.-N. 2015, \apj, 798, 84

\bibitem[{{Bai}(2016)}]{bai16}
---. 2016, \apj, 821, 80

\bibitem[{{Bai} \& {Stone}(2010{\natexlab{a}})}]{bai10}
{Bai}, X.-N., \& {Stone}, J.~M. 2010{\natexlab{a}}, \apj, 722, 1437

\bibitem[{{Bai} \& {Stone}(2010{\natexlab{b}})}]{bai10c}
---. 2010{\natexlab{b}}, \apjl, 722, L220

\bibitem[{{Balbus} \& {Hawley}(1991)}]{balbus91}
{Balbus}, S.~A., \& {Hawley}, J.~F. 1991, \apj, 376, 214

\bibitem[{Balsara {et~al.}(2009)Balsara, Tilley, Rettig, \&
  Brittain}]{balsara09}
Balsara, D.~S., Tilley, D.~A., Rettig, T., \& Brittain, S.~D. 2009, Monthly
  Notices of the Royal Astronomical Society, 397, 24.
\newblock \url{https://doi.org/10.1111/j.1365-2966.2009.14606.x}

\bibitem[{{Beck} {et~al.}(2010){Beck}, {Bary}, \& {McGregor}}]{beck10}
{Beck}, T.~L., {Bary}, J.~S., \& {McGregor}, P.~J. 2010, \apj, 722, 1360

\bibitem[{{Birnstiel} {et~al.}(2010){Birnstiel}, {Dullemond}, \&
  {Brauer}}]{birnstiel10}
{Birnstiel}, T., {Dullemond}, C.~P., \& {Brauer}, F. 2010, \aap, 513, A79

\bibitem[{{Blum}(2018)}]{blum18}
{Blum}, J. 2018, \ssr, 214, 52

\bibitem[{{Carrera} {et~al.}(2017){Carrera}, {Gorti}, {Johansen}, \&
  {Davies}}]{carrera17}
{Carrera}, D., {Gorti}, U., {Johansen}, A., \& {Davies}, M.~B. 2017, \apj, 839,
  16

\bibitem[{{Carrera} {et~al.}(2015){Carrera}, {Johansen}, \&
  {Davies}}]{carrera15}
{Carrera}, D., {Johansen}, A., \& {Davies}, M.~B. 2015, \aap, 579, A43

\bibitem[{{Chiang} \& {Youdin}(2010)}]{chiang10}
{Chiang}, E., \& {Youdin}, A.~N. 2010, Annual Review of Earth and Planetary
  Sciences, 38, 493

\bibitem[{{Dr{\c a}{\.z}kowska} \& {Dullemond}(2014)}]{drazkowska14}
{Dr{\c a}{\.z}kowska}, J., \& {Dullemond}, C.~P. 2014, \aap, 572, A78

\bibitem[{{Dr{\k{a}}{\.z}kowska} {et~al.}(2016){Dr{\k{a}}{\.z}kowska},
  {Alibert}, \& {Moore}}]{drazkowska16}
{Dr{\k{a}}{\.z}kowska}, J., {Alibert}, Y., \& {Moore}, B. 2016, \aap, 594, A105

\bibitem[{{Dubrulle} {et~al.}(1995){Dubrulle}, {Morfill}, \&
  {Sterzik}}]{dubruelle95}
{Dubrulle}, B., {Morfill}, G., \& {Sterzik}, M. 1995, \icarus, 114, 237

\bibitem[{Ercolano {et~al.}(2017)Ercolano, Jennings, Rosotti, \&
  Birnstiel}]{ercolano17}
Ercolano, B., Jennings, J., Rosotti, G., \& Birnstiel, T. 2017, Monthly Notices
  of the Royal Astronomical Society, 472, 4117

\bibitem[{{Flaherty} {et~al.}(2018){Flaherty}, {Hughes}, {Teague}, {Simon},
  {Andrews}, \& {Wilner}}]{flaherty18}
{Flaherty}, K.~M., {Hughes}, A.~M., {Teague}, R., {et~al.} 2018, \apj, 856, 117

\bibitem[{{Flaherty} {et~al.}(2017){Flaherty}, {Hughes}, {Rose}, {Simon}, {Qi},
  {Andrews}, {K{\'o}sp{\'a}l}, {Wilner}, {Chiang}, {Armitage}, \&
  {Bai}}]{flaherty17}
{Flaherty}, K.~M., {Hughes}, A.~M., {Rose}, S.~C., {et~al.} 2017, \apj, 843,
  150

\bibitem[{{Flock} {et~al.}(2017){Flock}, {Nelson}, {Turner}, {Bertrang},
  {Carrasco-Gonz{\'a}lez}, {Henning}, {Lyra}, \& {Teague}}]{flock17}
{Flock}, M., {Nelson}, R.~P., {Turner}, N.~J., {et~al.} 2017, \apj, 850, 131

\bibitem[{{Fromang} \& {Lesur}(2017)}]{fromang17}
{Fromang}, S., \& {Lesur}, G. 2017, arXiv e-prints, arXiv:1705.03319

\bibitem[{{Fromang} \& {Papaloizou}(2006)}]{fromang06}
{Fromang}, S., \& {Papaloizou}, J. 2006, \aap, 452, 751

\bibitem[{{Goldreich} \& {Lynden-Bell}(1965)}]{goldreich65}
{Goldreich}, P., \& {Lynden-Bell}, D. 1965, \mnras, 130, 125

\bibitem[{{Gole} {et~al.}(2020){Gole}, {Simon}, {Li}, {Youdin}, \&
  {Armitage}}]{gole20}
{Gole}, D.~A., {Simon}, J.~B., {Li}, R., {Youdin}, A.~N., \& {Armitage}, P.~J.
  2020, arXiv e-prints, arXiv:2001.10000

\bibitem[{{Jacquet} {et~al.}(2011){Jacquet}, {Balbus}, \& {Latter}}]{jacquet11}
{Jacquet}, E., {Balbus}, S., \& {Latter}, H. 2011, \mnras, 415, 3591

\bibitem[{{Johansen} {et~al.}(2014){Johansen}, {Blum}, {Tanaka}, {Ormel},
  {Bizzarro}, \& {Rickman}}]{johansen14}
{Johansen}, A., {Blum}, J., {Tanaka}, H., {et~al.} 2014, Protostars and Planets
  VI, 547

\bibitem[{{Johansen} {et~al.}(2011){Johansen}, {Klahr}, \&
  {Henning}}]{johansen11}
{Johansen}, A., {Klahr}, H., \& {Henning}, T. 2011, \aap, 529, A62

\bibitem[{{Johansen} \& {Youdin}(2007)}]{johansen07}
{Johansen}, A., \& {Youdin}, A. 2007, \apj, 662, 627

\bibitem[{{Johansen} {et~al.}(2009){Johansen}, {Youdin}, \& {Mac
  Low}}]{johansen09}
{Johansen}, A., {Youdin}, A., \& {Mac Low}, M.-M. 2009, \apjl, 704, L75

\bibitem[{{Klahr} \& {Hubbard}(2014)}]{klahr14}
{Klahr}, H., \& {Hubbard}, A. 2014, \apj, 788, 21

\bibitem[{{Klahr} {et~al.}(2018){Klahr}, {Pfeil}, \& {Schreiber}}]{klahr18}
{Klahr}, H., {Pfeil}, T., \& {Schreiber}, A. 2018, {Instabilities and Flow
  Structures in Protoplanetary Disks: Setting the Stage for Planetesimal
  Formation}, 138

\bibitem[{{Kowalik} {et~al.}(2013){Kowalik}, {Hanasz}, {W{\'o}lta{\'n}ski}, \&
  {Gawryszczak}}]{kowalik13}
{Kowalik}, K., {Hanasz}, M., {W{\'o}lta{\'n}ski}, D., \& {Gawryszczak}, A.
  2013, \mnras, 434, 1460

\bibitem[{{Krapp} {et~al.}(2019){Krapp}, {Ben{\'\i}tez-Llambay}, {Gressel}, \&
  {Pessah}}]{krapp19}
{Krapp}, L., {Ben{\'\i}tez-Llambay}, P., {Gressel}, O., \& {Pessah}, M.~E.
  2019, \apjl, 878, L30

\bibitem[{{Laibe} \& {Price}(2014)}]{laibe14}
{Laibe}, G., \& {Price}, D.~J. 2014, \mnras, 440, 2136

\bibitem[{{Latter} \& {Ogilvie}(2006)}]{latter06}
{Latter}, H.~N., \& {Ogilvie}, G.~I. 2006, \mnras, 372, 1829

\bibitem[{{Lesur} {et~al.}(2014){Lesur}, {Kunz}, \& {Fromang}}]{lesur14}
{Lesur}, G., {Kunz}, M.~W., \& {Fromang}, S. 2014, \aap, 566, A56

\bibitem[{{Lin}(2019)}]{lin19}
{Lin}, M.-K. 2019, \mnras, 485, 5221

\bibitem[{{Lin} \& {Kratter}(2016)}]{lin16}
{Lin}, M.-K., \& {Kratter}, K.~M. 2016, \apj, 824, 91

\bibitem[{{Lin} \& {Youdin}(2017)}]{lin17}
{Lin}, M.-K., \& {Youdin}, A.~N. 2017, \apj, 849, 129

\bibitem[{Lovascio \& Paardekooper(2019)}]{lovascio19}
Lovascio, F., \& Paardekooper, S.-J. 2019, MNRAS, 488, 5290

\bibitem[{{Lyra} \& {Umurhan}(2019)}]{lyra19}
{Lyra}, W., \& {Umurhan}, O.~M. 2019, \pasp, 131, 072001

\bibitem[{{Marcus} {et~al.}(2015){Marcus}, {Pei}, {Jiang}, {Barranco},
  {Hassanzadeh}, \& {Lecoanet}}]{marcus15}
{Marcus}, P.~S., {Pei}, S., {Jiang}, C.-H., {et~al.} 2015, \apj, 808, 87

\bibitem[{{Morfill} \& {Voelk}(1984)}]{morfill84}
{Morfill}, G.~E., \& {Voelk}, H.~J. 1984, \apj, 287, 371

\bibitem[{{Nelson} {et~al.}(2013){Nelson}, {Gressel}, \& {Umurhan}}]{nelson13}
{Nelson}, R.~P., {Gressel}, O., \& {Umurhan}, O.~M. 2013, \mnras, 435, 2610

\bibitem[{{Nesvorn{\'y}} {et~al.}(2019){Nesvorn{\'y}}, {Li}, {Youdin}, {Simon},
  \& {Grundy}}]{nesvorny19}
{Nesvorn{\'y}}, D., {Li}, R., {Youdin}, A.~N., {Simon}, J.~B., \& {Grundy},
  W.~M. 2019, Nature Astronomy, 364

\bibitem[{{Paardekooper} \& {Papaloizou}(2009)}]{paardekooper09d}
{Paardekooper}, S.~J., \& {Papaloizou}, J.~C.~B. 2009, \mnras, 394, 2283

\bibitem[{{Pinte} {et~al.}(2016){Pinte}, {Dent}, {M{\'e}nard}, {Hales}, {Hill},
  {Cortes}, \& {de Gregorio-Monsalvo}}]{pinte16}
{Pinte}, C., {Dent}, W.~R.~F., {M{\'e}nard}, F., {et~al.} 2016, \apj, 816, 25

\bibitem[{{Price} \& {Laibe}(2015)}]{price15}
{Price}, D.~J., \& {Laibe}, G. 2015, \mnras, 451, 813

\bibitem[{{Pringle}(1981)}]{pringle81}
{Pringle}, J.~E. 1981, \araa, 19, 137

\bibitem[{{Sch{\"a}fer} {et~al.}(2017){Sch{\"a}fer}, {Yang}, \&
  {Johansen}}]{schafer17}
{Sch{\"a}fer}, U., {Yang}, C.-C., \& {Johansen}, A. 2017, \aap, 597, A69

\bibitem[{{Shakura} \& {Sunyaev}(1973)}]{shakura73}
{Shakura}, N.~I., \& {Sunyaev}, R.~A. 1973, \aap, 24, 337

\bibitem[{{Shi} \& {Chiang}(2013)}]{shi13}
{Shi}, J.-M., \& {Chiang}, E. 2013, \apj, 764, 20

\bibitem[{{Simon} {et~al.}(2016){Simon}, {Armitage}, {Li}, \&
  {Youdin}}]{simon16}
{Simon}, J.~B., {Armitage}, P.~J., {Li}, R., \& {Youdin}, A.~N. 2016, \apj,
  822, 55

\bibitem[{{Simon} {et~al.}(2017){Simon}, {Armitage}, {Youdin}, \&
  {Li}}]{simon17}
{Simon}, J.~B., {Armitage}, P.~J., {Youdin}, A.~N., \& {Li}, R. 2017, \apjl,
  847, L12

\bibitem[{Simon {et~al.}(2018)Simon, Bai, Flaherty, \& Hughes}]{simon18}
Simon, J.~B., Bai, X.-N., Flaherty, K.~M., \& Hughes, A.~M. 2018, The
  Astrophysical Journal, 865, 10

\bibitem[{{Squire} \& {Hopkins}(2018)}]{squire17}
{Squire}, J., \& {Hopkins}, P.~F. 2018, \mnras, 477, 5011

\bibitem[{{Stoll} \& {Kley}(2016)}]{stoll16}
{Stoll}, M. H.~R., \& {Kley}, W. 2016, \aap, 594, A57

\bibitem[{{Umurhan} {et~al.}(2019){Umurhan}, {Estrada}, \& {Cuzzi}}]{umurhan19}
{Umurhan}, O.~M., {Estrada}, P.~R., \& {Cuzzi}, J.~N. 2019, arXiv e-prints,
  arXiv:1906.05371

\bibitem[{{Weidenschilling}(1977)}]{weiden77}
{Weidenschilling}, S.~J. 1977, \mnras, 180, 57

\bibitem[{{Whipple}(1972)}]{whipple72}
{Whipple}, F.~L. 1972, in From Plasma to Planet, ed. A.~{Elvius}, 211

\bibitem[{{Yang} \& {Johansen}(2014)}]{yang14}
{Yang}, C.-C., \& {Johansen}, A. 2014, \apj, 792, 86

\bibitem[{{Yang} {et~al.}(2018){Yang}, {Mac Low}, \& {Johansen}}]{yang18}
{Yang}, C.-C., {Mac Low}, M.-M., \& {Johansen}, A. 2018, \apj, 868, 27

\bibitem[{{Youdin} \& {Johansen}(2007)}]{youdin07b}
{Youdin}, A., \& {Johansen}, A. 2007, \apj, 662, 613

\bibitem[{{Youdin}(2011)}]{youdin11}
{Youdin}, A.~N. 2011, \apj, 731, 99

\bibitem[{{Youdin} \& {Goodman}(2005)}]{youdin05a}
{Youdin}, A.~N., \& {Goodman}, J. 2005, \apj, 620, 459

\bibitem[{{Youdin} \& {Lithwick}(2007)}]{youdin07}
{Youdin}, A.~N., \& {Lithwick}, Y. 2007, \icarus, 192, 588

\end{thebibliography}
